\pdfoutput=1
\documentclass[11pt]{article}

\usepackage[]{ACL2023}

\usepackage{times}
\usepackage{latexsym}
\usepackage{amsmath}
\usepackage{adjustbox}
\usepackage{multirow}
\usepackage{subcaption}
\usepackage{caption}
\usepackage{balance} 
\usepackage{graphicx}
\usepackage{makecell}
\usepackage{amssymb}
\usepackage{anyfontsize}
\usepackage{arydshln}
\usepackage[T1]{fontenc}
\usepackage{microtype}

\usepackage[utf8]{inputenc}
\usepackage{inconsolata}

\title{Interpretable Multimodal Misinformation Detection with Logic Reasoning}

 \author{Hui Liu\textsuperscript{1} \quad \quad  \quad \quad 
  Wenya Wang\textsuperscript{2,3} \quad \quad \quad \quad  Haoliang Li\textsuperscript{1}\\
  \textsuperscript{1}City University of Hong Kong \\
    \textsuperscript{2}Nanyang Technological University\\
  \textsuperscript{3}University of Washington \\ 
\text{liuhui3-c@my.cityu.edu.hk}, \text{wangwy@ntu.edu.sg}, \text{haoliang.li@cityu.edu.hk}
}
\begin{document}
\maketitle

\begin{abstract}

Multimodal misinformation on online social platforms is becoming a critical concern due to increasing credibility and easier dissemination brought by multimedia content, compared to traditional text-only information. While existing multimodal detection approaches have achieved high performance, the lack of interpretability hinders these systems' reliability and practical deployment. Inspired by Neural-Symbolic AI which combines the learning ability of neural networks with the explainability of symbolic learning, we propose a novel logic-based neural model for multimodal misinformation detection which integrates interpretable logic clauses to express the reasoning process of the target task. To make learning effective, we parameterize symbolic logical elements using neural representations, which facilitate the automatic generation and evaluation of meaningful logic clauses. Additionally, to make our framework generalizable across diverse misinformation sources, we introduce five meta-predicates that can be instantiated with different correlations. Results on three public datasets (Twitter, Weibo, and Sarcasm) demonstrate the feasibility and versatility of our model. The implementation of our work can be found in this link \footnote{\url{https://github.com/less-and-less-bugs/LogicMD}}.
\end{abstract}

\section{Introduction}
Misinformation refers to incorrect or misleading information\footnote{\url{https://www.merriam-webster.com/dictionary/misinformation}} which includes fake news, rumors, satire, etc. The enormous amount of misinformation emerged on online social platforms is attributed to users' reliability on the
information provided by the internet and the inability to discern fact from fiction \cite{spinney2017facebook}. Moreover, widespread misinformation can have negative consequences for both societies and individuals. Therefore, there is an urgent need to identify misinformation automatically. While numerous posts are in multimodal style (i.e., text and image) on social media, this work concentrates on multimodal misinformation detection. 

\begin{figure}[t]
\centering
\begin{subfigure}{0.235\textwidth}
\centering
\includegraphics[width=\linewidth]{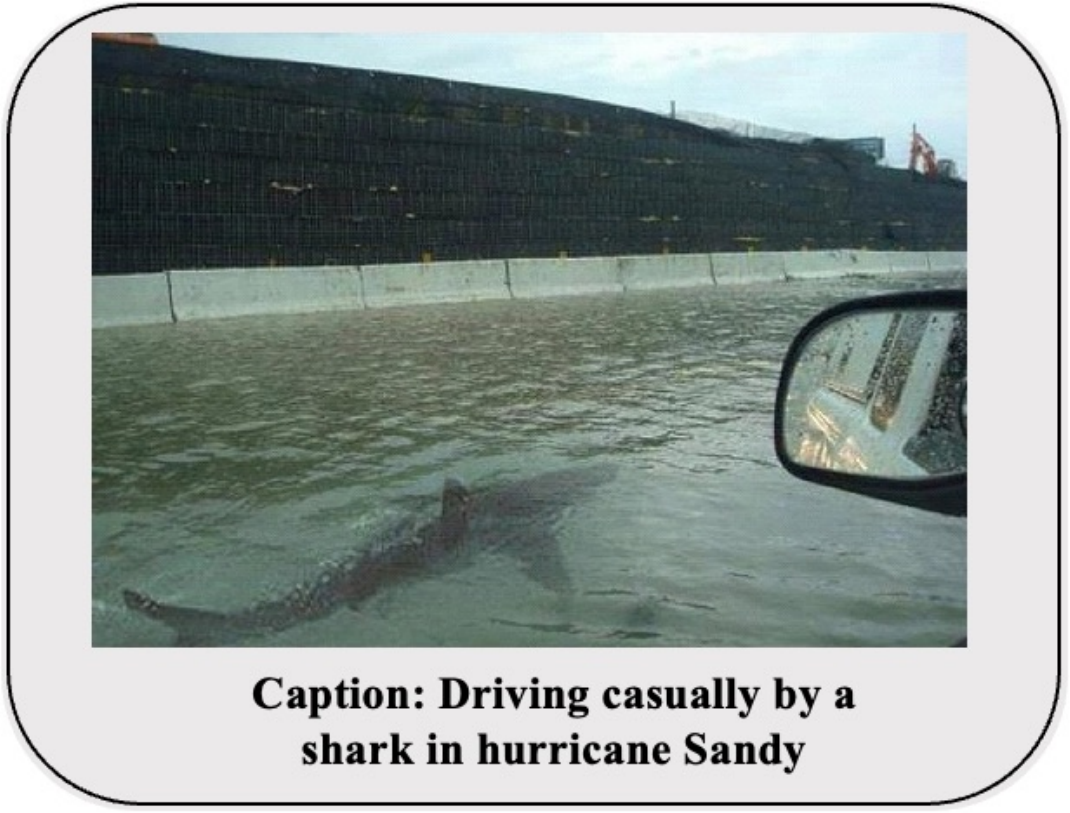}
\caption{Original Rumor}
\label{MCA}
\end{subfigure}
\hfill
\begin{subfigure}{0.235\textwidth}
\includegraphics[width=\linewidth]{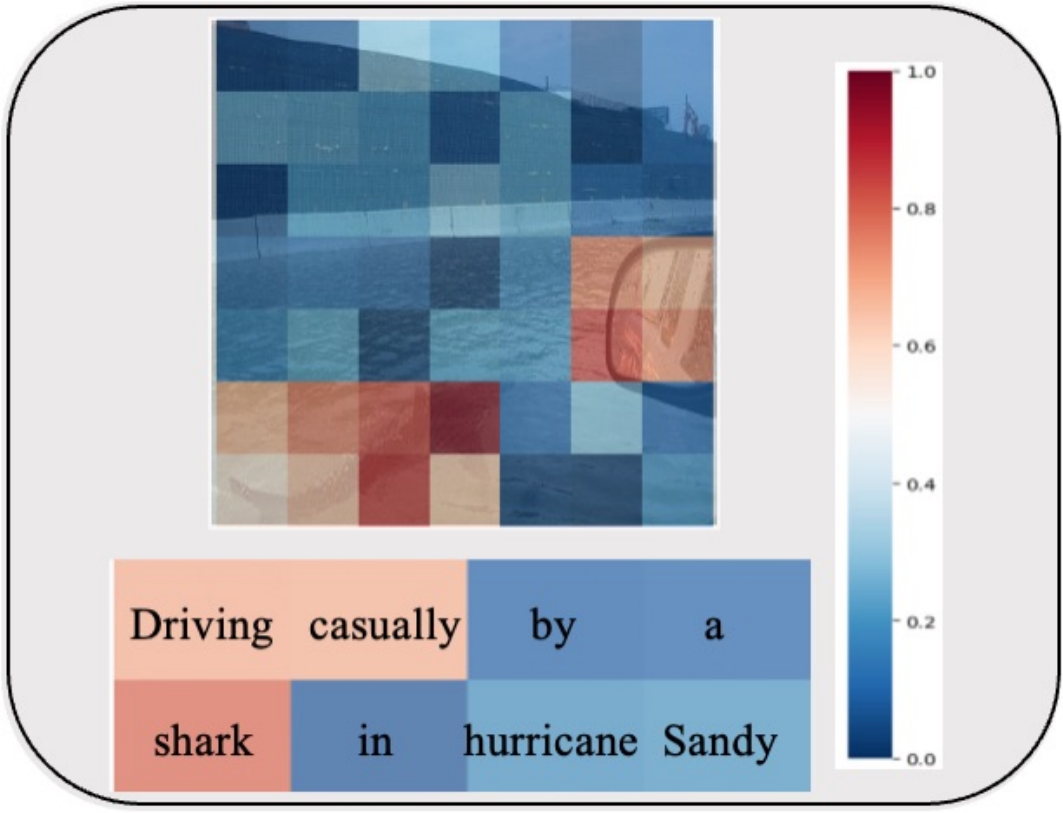} 
\centering
\caption{Attention Map}
\label{gat}
\end{subfigure}
\begin{subfigure}{0.235\textwidth}
\includegraphics[width=\linewidth]{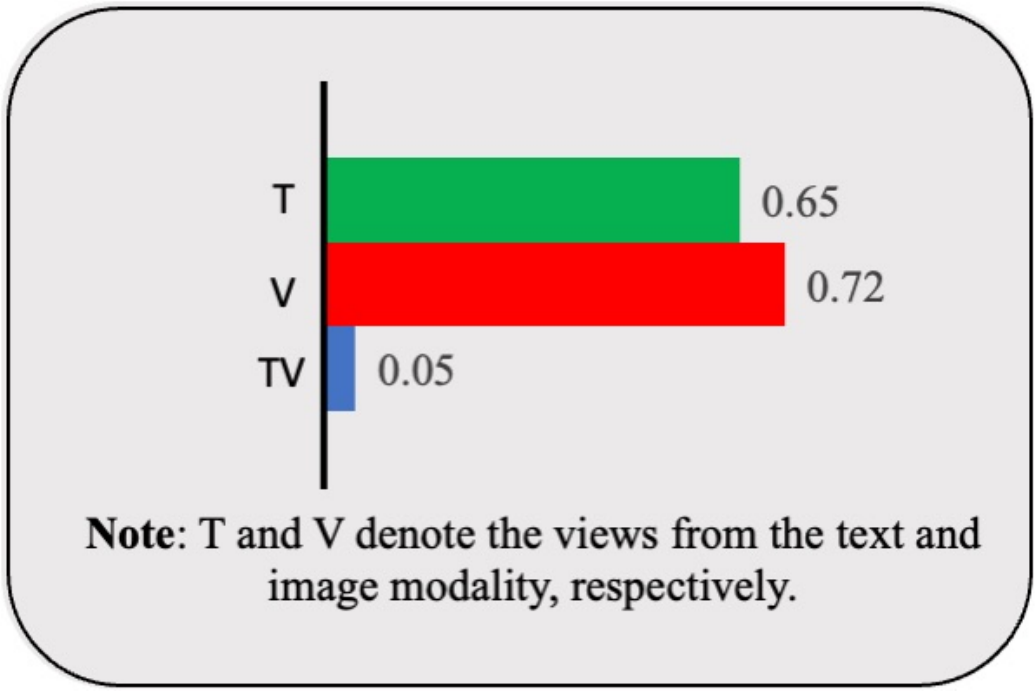} 
\centering
\caption{Multi-View}
\label{gat}
\end{subfigure}
\hfill
\begin{subfigure}{0.235\textwidth}
\includegraphics[width=\linewidth]{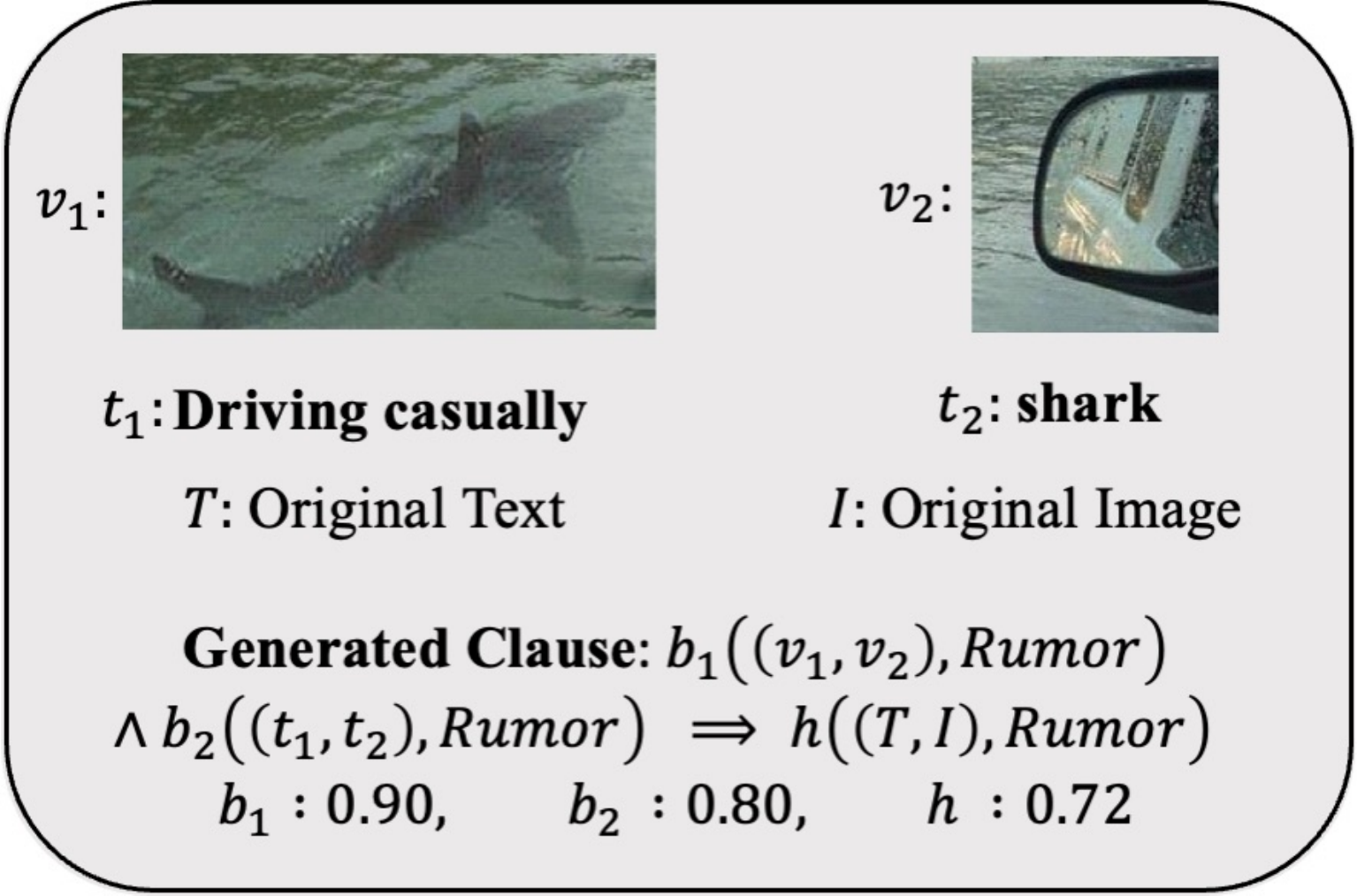} 
\centering
\caption{Neural-Symbolic}
\label{ns}
\end{subfigure}
\caption{Examples of explanations generated by attention map, multi-view, and our proposed Neural-Symbolic-based method for a rumor sample in Twitter dataset. For (c) and (d), a higher value indicates a higher probability of being detected as a rumor.}
\vspace{-10pt}
\label{examplesofexplain}
\end{figure}

Multimodal approaches, which either fuse text and image features \cite{eann18,MVAE19, MVNN21, www22CAFE} or investigate discrepancies between the two modalities  \cite{vt1,  vt2}, have been used for misinformation detection with some success. However, these methods often lack interpretability because of the black-box nature of the neural network. Some frameworks have been proposed to solve this challenge. As depicted in Fig.~\ref{examplesofexplain}, methods based on attention maps, such as those outlined in \citep{IncrossMGsacl22} and \citep{HCMemnlp22}, have been employed to identify highly correlated text or image content  (referred to here as "where") according to attention weights, while multi-view based methods, such as those described in \citep{t2TKDE22} and \citep{BMRAAAI23}, have been utilized to highlight the most contributive perspectives\footnote{Perspective is defined as a particular aspect to identify misinformation. In our work, it involves different types of assembly of different modalities, following a popular classification method of existing misinformation detection approaches \citep{fakenewssurveycoling22}.} (referred to here as "how"). However, the explainability of the fusion of such attention or views has yet to be fully established \citep{ICMLattention2022}, and these methods cannot concurrently illustrate both the ``where" and ``how" of the reasoning process. Such interpretability is crucial for ensuring trust, reliability, and adoption of deep learning systems in real-world applications \cite{explainableai, sun2021personalized,cui2022privacy}, particularly when it comes to detecting misinformation \cite{explain3kdd19}.

% \footnote{\url{https://www.snopes.com/fact-check/shark-street-hurricane}}
 To address the aforementioned limitations, owing to Neural-Symbolic learning \citep{nsaiijcai20, nasisurvey2}, we propose to  incorporate logic reasoning into the misinformation detection framework to derive human-readable clauses. As shown in Fig.~\ref{ns}, the clause $b_1((v_1,v_2), Rumor) \land b_2((t_1,t_2), Rumor) \Rightarrow h((T,I)), Rumor)$ is induced from the text-image pair where constants $v_1$, $v_2$, $t_1$, $t_2$ are crucial visual patches and textual tokens for predication, corresponding to "where". Body predicates $b_1$ and $b_2$ indicate relationships between patches and tokens for misinformation identification, corresponding to "how". We propose to automatically learn these logic clauses which explicitly express evident features and their interactions to promote interpretability and improve the final performance, which has not been explored by previous work. 
 
 However, given the intrinsic complexity and diversity of multimodal context, it is hard to explicitly predefine the exact relationships as logic predicates. To this end, we introduce five general perspectives relevant to the task of misinformation detection as meta-predicates for clause formulation. These perspectives include suspicious atomic textual content, visual content, relationships between text tokens, visual patches and both modalities. Each meta-predicate can be instantiated with different correlations between contents of the text-image pair and target labels (e.g., $(t_1, t_2)$ and $Rumor$ in Fig.~\ref{ns}), aiming to cover a wide range of aspects leading to misinformation. For instance, the fifth perspective implicates exploiting cross-modal contents to debunk misinformation while cross-modal ambiguity learning \cite{www22CAFE}, inconsistency between news contents and background knowledge \cite{abdelnabi2022open} and entities misalignment \cite{vt1} are candidate correlations to achieve this goal.
 
 Building upon these definitions, we propose a logic-based multimodal misinformation detection model (\textbf{LogicDM}). LogicDM first extracts embeddings for text tokens and image patches using corresponding encoders and then generates cross-modal object embeddings for different predicates using a multi-layer graph convolutional network (GCN). We then propose to parameterize meta-predicates by weighing the importance of each correlation. When combined with different object constants, these meta-predicates are softly selected to produce interpretable logic clauses defining the target predicate. The whole framework can be trained end-to-end with differentiable logic operators and probabilistic logic evaluations. To summarize, the contributions of this work include: 1) We propose an explainable neural-symbolic approach capable of automatically generating logic clauses instantiated with multimodal objects via differentiable neural components. 2) We define five meta-predicates building upon existing misinformation detection perspectives and introduce an adaptive mechanism to represent these predicates using soft selections over multiple pre-defined correlations. 3) We provide comprehensive evaluations of our model on three benchmark datasets. 

\section{Related Work}
\subsection{Misinformation Detection}
\label{fivepredicates}

Misinformation detection has gained significant attention in recent years due to the proliferation of content on online social media \cite{fakenewssurveycoling22}. To identify misinformation, the text modality can be used with clues such as semantics \cite{t2TKDE22, t6}, writing style \cite{t4}, emotion \cite{t2TKDE22}, special word usage \cite{t3SIGIR22}, and punctuation \cite{t1, t5}. In addition, image features can help detect misinformation, with fake and real news often having distinct image distribution patterns, including differences in image semantics and compression trace \cite{weiboset, v1}. Intra-modal inconsistency and incongruity within the text or image \cite{tt1, vv1} can also serve as indicators of misinformation. Cross-modal interaction and fusion, used by many recent multimodality-based methods, can assist in detecting misinformation. For example, \cite{vt1,vt2} compared the characteristics of entities across the textual and visual modalities, while \citet{BMRAAAI23} measured cross-modal inconsistency through Kullback-Leibler divergence between unimodal distributions.

\subsection{Neural-Symbolic Reasoning}

Deep learning has achieved impressive results, but its limitations in interpretability and logical reasoning have been noted by \citep{nasisurvey2}. To address these limitations, the integration of symbolic reasoning and neural networks, known as Neural-Symbolic AI, has gained attention as a potential solution \cite{nsaiijcai20}. One approach enhances neural networks with structured logic rules, such as first-order logic, that act as external constraints during model training \cite{hu2016harnessing, deeperprolognips18, wenyacl21, logicfactverification}. The other approach, Inductive Logic Programming (ILP), aims to automatically construct first-order logic rules from noisy data \cite{ilpsurvey1}. There have been various proposed ILP architectures, including NeuralLP \cite{yang2017nlp}, LNN \cite{LNNplus}, $\delta$ILP \cite{evans2018learning}, and RNNLogic \cite{rnnlogic}. ILP has been applied in a range of areas including knowledge-base completion \cite{rnnlogic}, question answering \cite{logicthrmacl22}, and multi-hop reading comprehension \cite{wenyaacl22multihopreasoning}. However, multimodal misinformation detection, unlike these previous applications, faces the challenge of lacking well-defined predicates and constants due to the unstructured and modality-different text-image input.

\section{Preliminaries}
\subsection{Task Definition}
In this paper, we aim to address the problem of multimodal misinformation detection. Given a text-image pair $(T, I)$, we seek to predict its label. To incorporate logic reasoning into the neural network, we define a candidate label set $\mathcal{Y} = \{\textrm{NonRumor}, \textrm{Rumor}\}$ for rumor detection task while  $\mathcal{Y} = \{\textrm{NonSarcasm}, \textrm{Sarcasm}\}$ for sarcasm detection task. We also define a 2-ary predicate $h$ that takes as input a text-image pair and a label, with the implicit meaning that the text-image pair satisfies the label. Our goal can then be reformulated as selecting a label $y \in \mathcal{Y}$ such that $h((T, I), y)$ holds. It is worth noting that this definition allows for the extension of our framework to multi-class classification tasks by increasing the size of the set of labels $\mathcal{Y}$.

\subsection{Inductive logic programming}
\label{ilpsubsection}
To address the interpretability challenge in misinformation detection, we propose a framework that induces rules or clauses of the form $b_1 \land \ldots \land b_q \Rightarrow h$, where $b_1, \ldots, b_q$ are predicates in the body, $h$ is the head predicate, and $\land$ denotes the conjunction operation. The body predicates are 2-ary, defined over object variable $O$ (i.e., combinations of text tokens, image patches, or both) and label variable $Y$ (i.e., labels in the set $\mathcal{Y}$). These predicates with associated variables, such as $b(O, Y)$, are referred to as logic atoms. By instantiating variables in body atoms with constants (e.g., $b(o, y)$, where $o$ is an object and $y$ is a label), we can obtain truth values of these body atoms and subsequently derive the value of the head atom $h((T, I), y)$ using logic operators (e.g., conjunction $\land$ and disjunction $\lor$), where the truth value indicates the probability of the atom or clause being true and is in the range of 0 to 1, denoted as $\mu(\cdot) \in \left[0, 1 \right]$. 

\section{Methodology}
\begin{figure}[t]
\centering
\includegraphics[width=1\linewidth]{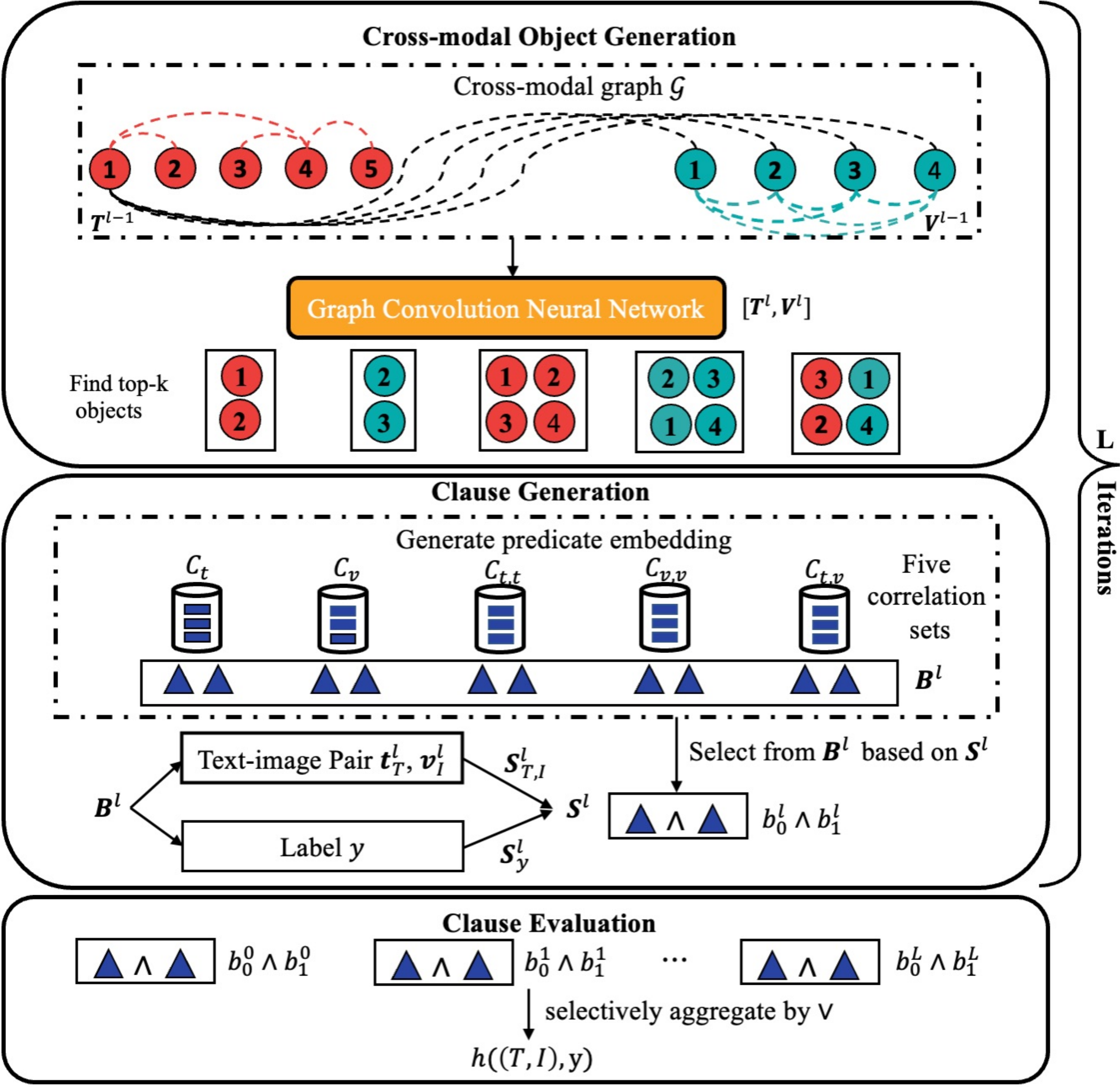}
\caption{The core architecture of the proposed interpretable multimodal misinformation detection framework based on logic reasoning (\textbf{LogicDM}). Textual nodes are fully connected to visual nodes but we only visualize edges between one textual node and visual nodes for ease of illustration.}
\label{overview}
\vspace{-10pt}
\end{figure}
This section introduces the proposed logic-based multimodal misinformation detection model (\textbf{LogicDM}), which offers a more explicit reasoning process and better performance than existing approaches. The model consists of four main components: Feature Extraction, Cross-modal Object Generation, Clause Generation, and Clause Evaluation. Feature Extraction generates representations for text tokens and image patches using encoders. Cross-modal Object Generation constructs a cross-modal graph and applies a multi-layer graph convolutional neural network to generate multi-grained representations that constitute cross-modal objects as logic constants. Clause Generation produces dynamic embeddings for predicates (see Table 1) by weighing the importance of different correlations and considers the logic relationship among all predicates to adaptively derive probable logic clauses. These clauses, when instantiated with object constants, can be evaluated to determine the truth value as Clause Evaluation. The overview of this model is shown in Fig.~\ref{overview} and a running example is depicted in Fig.~ \ref{runningsample}. 

\subsection{Feature Extraction}
Given text-image pair $(T, I)$ as input, we first tokenize $T$ into $m$ tokens, denoted as $X_T = \{w_1, w_2, \ldots, w_m\}$. Then we use BERT \citep{bert}  with a one-layer LSTM \citep{lstm} as the textual encoder to obtain $d$-dimension representations for all tokens in $X_T $, given as $\mathbf{T}=\left[\mathbf{t}_1,  \mathbf{t}_2, \ldots, \mathbf{t}_{m} \right]$, where $\mathbf{T}\in\mathbb{R}^{m \times d}$.

For image modality, we first resize the image to the size $224 \times 224$ and divide each image into $r=z^2$ patches, where the size of each patch is $224/z \times 224/z$. Similar to text modality, these patches are reshaped to a sequence, denoted as $X_I = \{p_1, p_2, \ldots, p_r\}$. Then we exploit the pre-trained visual backbone neural network (e.g., ResNet34 \citep{resnet34} and ViT \citep{iclr21vit}) to extract visual features and map these features to $d$-dimension using a two-layer MLP as  $\mathbf{V} = \left [\mathbf{v}_1, \mathbf{v}_2,\ldots, \mathbf{v}_{r} \right]$, where $\mathbf{V} \in \mathbb{R}^{r\times d}$.

\subsection{Cross-modal Object Generation}
Cross-modal Object Generation aims to produce representations for constants (e.g., $(v_1, v_2)$, $(t_1, t_2)$ in Fig.~\ref{examplesofexplain}) to instantiate logic clauses. Different from the common definition of constants as single objects (in images or texts), we define constants according to our newly introduced meta-predicates. Specifically, we define meta-predicates as higher-level perspectives pertinent to discriminating  misinformation. For this task, we use five meta-predicates, namely $b_t$ for single-token perspective, $b_{v}$ for single-image-patch perspective, $b_{t,t}$ for intra-text interactions, $b_{v,v}$ for intra-image interactions and $b_{t,v}$ for inter-modal interactions. The detailed explanations are shown in Table~\ref{predicate}. The constants for these meta-predicates include a single token $t_i$, a single image patch $v_i$, a pair of tokens $(t_i,t_j)$, a pair of image patches $(v_i, v_j)$, and a pair consisting of both modalities $(t_i, v_j)$. The representations, denoted by $\mathbf{o}$, for these constants are computed according to the formula in Table~\ref{predicate} and will be illustrated next.

\begin{table*}
\centering
    \begin{adjustbox}{max width=0.9\linewidth}
        \begin{tabular}{c|c|c}
        \hline
        Logic Atom & Predicate Meaning & Formula of Objects \\
        \hline
        $b_t(t, y)$ & token $t$ is related to label $y$ & $\mathbf{o}_t = \mathbf{t}\mathbf{W}_t$, $\mathbf{W}_t \in \mathbb{R}^{d \times d}$ \\
        \hline
        $b_{v}(v, y)$ & image patch $v$ is related to label $y$ & $\mathbf{o}_v=\mathbf{v} \mathbf{W}_v$, $\mathbf{W}_v \in \mathbb{R}^{d \times d}$ \\
        \hline
        $b_{t, t}((t_i,t_j), y)$ & the pair of tokens $(t_i,t_j)$ is related to label $y$ & $\mathbf{o}_{t_i, t_j}=[\mathbf{t}_i, \mathbf{t}_j, \mathbf{t}_i-\mathbf{t}_j, \mathbf{t}_i 
          \circ\mathbf{t}_j] \textbf{W}_{t,t}$, $\textbf{W}_{t, t} \in \mathbb{R}^{4d \times d}$ 
          \\
        \hline
        $b_{v, v}((v_i,v_j), y)$ & the pair of patches $(v_i,v_j)$ is related to label $y$ & $\mathbf{o}_{v_i, v_j}=[\mathbf{v}_i, \mathbf{v}_j, \mathbf{v}_i-\mathbf{v}_j, \mathbf{v}_i 
          \circ \mathbf{v}_j ]\textbf{W}_{v,v} $, $\textbf{W}_{v,v} \in \mathbb{R}^{4d \times d}$   \\
        \hline
        $b_{t, v}((t_i,v_j), y)$ & the pair of token and patch $(t_i,v_j)$ is related to label $y$  & $\mathbf{o}_{t_i, v_j} = [\mathbf{t}_i, \mathbf{v}_j, \mathbf{t}_i-\mathbf{v}_j, \mathbf{t}_i 
          \circ\mathbf{v}_j] \textbf{W}_{t,v} $, $\textbf{W}_{t, v} \in \mathbb{R}^{4d \times d}$\\
        \hline
        \end{tabular}
    \end{adjustbox}
\caption{The meaning of proposed five meta-predicates and formulas to produce cross-modal objects for each predicate. $\mathbf{t}^l \in \mathbb{R}^d$ and $\mathbf{v}^l \in \mathbb{R}^d$ denote textual and visual features obtained in the $l$-th iteration of GCN, and the subscripts $i$ and $j$ represents two different features. The bold symbol $\mathbf{o} \in \mathbb{R}^{d}$ represents the embedding of corresponding constant. And $\textbf{W}_{t}$, $\textbf{W}_{v}$, $\textbf{W}_{t, t}$, $\textbf{W}_{v, v}$ and $\textbf{W}_{t, v}$ are trainable parameters.}
\label{predicate}
\vspace{-10pt}
\end{table*}

The atoms, defined in Table~\ref{predicate}, necessitate disparate uni-modal and cross-modal inputs, thus, requiring our model to capture intricate intra-modal and inter-modal representations concurrently. Inspired by recent work on multimodal task \cite{IncrossMGsacl22, crossmodagraph}, we propose to construct a cross-modal graph $\mathcal{G}$ for $(T, I)$ to leverage the relations among text tokens $X_{T}$, image patches $X_I$ as well as those units between both modalities for computing representations of cross-modal constants.

Concretely, we take textual tokens $X_{T}$ and visual patches $X_I$ as nodes of graph $\mathcal{G}$, i.e., the node matrix is the concatenation of $X_{T}$ and $X_I$, denoted as $\left[ X_{T}, X_I \right]$ and the initial node embedding matrix is the concatenation of text-modality and image-modality representations, denoted as $\mathbf{H} = \left[ \mathbf{T}, \mathbf{V} \right]$, where $\mathbf{H} \in \mathbb{R}^{(m+r)\times d}$. For edges, the semantic dependencies among textual tokens are first extracted by Spacy\footnote{\url{https://spacy.io/}}. And if there exits a dependency between any two tokens, there will be an edge between them in $\mathcal{G}$. Then visual patches are connected according to their geometrical adjacency in the image, following \cite{HCMemnlp22}. Additionally, we assume the text nodes and visual nodes are fully connected to each other to increase interactions between two modalities, thus reducing the modality gap. Finally, the adjacency matrix $\mathbf{A} \in \mathbb{R}^{(m+r)\times (m+r)}$ can be represented as
\begin{equation}
\small
\mathbf{A}_{ij} = \begin{cases} 
1,\! &\! if \,i, j \leq m \textrm{ and }\textrm{a dependency exists in } w_i,w_j\\
1,\! &\! if \,i\leq m, j>m\, \textrm{or }i> m, j\leq m  \\
1,\! &\! if\, i,j>m\textrm{ and } p_{i-m}\textrm{, } p_{j-m} \textrm{ are adjacent},
\label{adja}
\end{cases}
\end{equation}
where $p_{i-m}$ and $p_{j-m}$ are determined as adjacent when $\small{\left\vert(i-m) \textrm{ mod }z-(j-m)\textrm{ mod }z \right\vert \leq 1}$ and $\small{\left\vert (i-m)/z-(j-m)/z \right\vert \leq 1}$. Subsequently, a $L$-layer GCN \cite{gcniclr17} is used to update each node embedding after fusing the information from its neighbor nodes via $\small{ \mathbf{H}^l = \textrm{ReLU}(\tilde{\mathbf{A}}\mathbf{H}^{l-1}\mathbf{W}^l)}$, where $l \in \{0, 1, \ldots, L\}$ represents the $l$-th iteration of GCN, $\tilde{\mathbf{A}} = \mathbf{D}^{-\frac{1}{2}}\mathbf{A} \mathbf{D}^{-\frac{1}{2}}$, 
$\mathbf{D}$ is the degree matrix of $\mathbf{A}$,  
% where $\mathbf{D}_{ii} = \sum_{j}\mathbf{A}_{i,j}$, 
and $\mathbf{W}^l \in \mathbb{R}^{d\times d}$ is a layer-specific trainable weight matrix. $\mathbf{H}^l \in \mathbb{R}^{(m+r)\times d}$ denotes the output of $l$-th GCN where $\mathbf{H}^l = {\left[ \mathbf{T}^l, \mathbf{V}^l \right]}$ and $\mathbf{H}^0 = \mathbf{H}$. Especially,  $\mathbf{T}^l \in\mathbb{R}^{m \times d}$  and $\mathbf{V}^l \in\mathbb{R}^{r \times d}$ are updated textual and visual representations at the $l$-th layer.

With $\mathbf{T}^l$ and $\mathbf{V}^l$, we compute representations of the cross-modal objects $\mathbf{O}^l_t \in \mathbb{R}^{m \times d}$, $\mathbf{O}^l_v \in \mathbb{R}^{r \times d}$, $\mathbf{O}^l_{t,t} \in \mathbb{R}^{(m\times m) \times d}$,  $\mathbf{O}^l_{v,v} \in \mathbb{R}^{(r \times r) \times d}$ and $\mathbf{O}^l_{t,v} \in \mathbb{R}^{(m\times r)\times d}$ as constants for those meta-predicates, according to formulas in Table~\ref{predicate}. In subsequent illustrations, we omit the layer index $l$ for ease of illustration. Intuitively, different objects have different importance for multimodal misinformation detection task. As such, we feed the embedding of each object to a separate MLP (one linear layer with a ReLU as the activation function) to compute its importance score corresponding to a specific meta-predicate. The embedding of Then $k$ objects are chosen for each meta-predicate based on their importance scores for clause generations and evaluations and the their embedding will be  . We denote their representations as $\hat{\mathbf{O}}_t$, $\hat{\mathbf{O}}_v$, $\hat{\mathbf{O}}_{t,t}$, $\hat{\mathbf{O}}_{v,v}$ and $\hat{\mathbf{O}}_{t,v}$, each of which belongs to $ \mathbb{R}^{k\times d}$.

\subsection{Clause Generation}
\label{Clause-Generation}
In Clause Generation, we derive logic clauses consisting of meta-predicates that deduce the head atom $h((T, I), y)$, e.g., $b_v(v,y) \wedge b_t(t,y) \Rightarrow h((T,I),y)$. For each meta-predicate, we pre-define a set of $g$ fine-grained correlations (parameterized with embeddings) between objects and labels, denoted by $\mathbf{C} \in \mathbb{R}^{g\times d}$ (i.e., $\mathbf{C}_t$, $\mathbf{C}_v$, $\mathbf{C}_{t,t}$, $\mathbf{C}_{v,v}$, $\mathbf{C}_{t,v}$ corresponding to $b_t$, $b_v$, $b_{t,t}$, $b_{v,v}$ and $b_{t,v}$, respectively). 
For example, $\mathbf{C}_{t}$ stores $g$ correlations between text tokens and labels relevant to meta-predicate $b_t(t, y)$. 
These correlations can be flexibly combined to form an embedding for each meta-predicate with different instantiations.

Concretely, taking meta-predicate $b_t(t,y)$ as an example, the embedding $\mathbf{B}_t$ for $b_t(t,y)$ with all instantiations $t$ (i.e., $\hat{\textbf{O}}_t$) is computed as 
\begin{equation}
\label{o-predicate}
\small
 \mathbf{B}_{t} = \textrm{sparsemax}([\hat{\textbf{O}}_t, \mathbf{y}]\textbf{W}^e_t\mathbf{C}_{t}^\top)\mathbf{C}_{t}. 
\end{equation}
Here  $\mathbf{B}_{t} \in \mathbb{R}^{k \times d}$ consists of $k$ embeddings corresponding to $k$ different objects extracted in $\hat{\mathbf{O}}_t$. $\mathbf{y}$ is the $d$-dimension embedding of label $y$ and is broadcasted to $k\times d$ for concatenation. $\textbf{W}^e_t \in \mathbb{R}^{2d \times d}$ is a learnable matrix. In addition, we utilize sparsemax, a sparse version of softmax, to select only a small number of correlations, which has been proven effective in multi-label classification tasks \cite{sparsemax}. The intuition of Eq.~\ref{o-predicate} is to softly select correlations to form the meta-predicate embedding when the input constants are $t$ and $y$. By adapting Eq.~\ref{o-predicate} to other meta-predicates, we obtain a complete set of predicate embeddings $\mathbf{B} \in \mathbb{R}^{5k\times d}$ where $\mathbf{B} = [\mathbf{B}_{t}, \mathbf{B}_{v}, \mathbf{B}_{t,t}, \mathbf{B}_{v,v},\mathbf{B}_{t,v}]$.

Furthermore, we obtain the embedding of the entire text input $\mathbf{t}_T \in \mathbb{R}^d$ and image $\mathbf{v}_I \in \mathbb{R}^d$ via weighed summations of all tokens and patches, respectively: $\mathbf{t}_T={\mathbf{T}}^\top \textrm{softmax}(\mathbf{T}\mathbf{W}_T)$ and $\mathbf{v}_I={\mathbf{V}}^\top\textrm{softmax}(\mathbf{V} \mathbf{W}_I)$, where $\mathbf{W}_T \in\mathbb{R}^{d \times 1 }$ and $\mathbf{W}_I \in\mathbb{R}^{d \times 1 }$ are trainable parameters to compute importance scores of tokens and patches. 

To generate valid clauses, given the predicate embeddings $\mathbf{B}$, textual representation $\mathbf{t}_T$ and image representation $\mathbf{v}_I$, we 
use two sparse attention networks to select relevant predicates pertinent to the image-text input, as well as the given label, to form the body of a clause. Formally, we have two attention scores $\mathbf{S}_{T,I}$ and $\mathbf{S}_{y}$ indicative of the input text-image pair and label respectively, given as
\begin{equation}
\small
    \begin{aligned}
    \mathbf{S}_{T,I} &=  \textrm{sparsemax}(\mathbf{B} \mathbf{W}_{T,I}[ \mathbf{t}_{T}, \mathbf{v}_{I}]), \\
    \mathbf{S}_{y} &=  \textrm{sparsemax}([\mathbf{B}, \mathbf{y}, \mathbf{B}-\mathbf{y}, \mathbf{B} \circ \mathbf{y}] \mathbf{W}_y),
    \end{aligned}
\end{equation}
% \small{
%     \begin{eqnarray}
%         \mathbf{S}_{T,I} &=&  \textrm{sparsemax}( \hat{\mathbf{B}} \mathbf{W}_{T,I}[ \mathbf{t}_{T}, \mathbf{v}_{I}]), \\
%         \mathbf{S}_{y} &=&  \textrm{sparsemax}([\hat{\mathbf{B}}, \mathbf{y}, \hat{\mathbf{B}}-\mathbf{y}, \hat{\mathbf{B}}\circ \mathbf{y}] \mathbf{W}_y),
%     \end{eqnarray}
% }
where $\mathbf{W}_{T,I} \in \mathbb{R}^{d\times2d}$ and $\mathbf{W}_{y} \in \mathbb{R}^{4d \times 1}$ are learnable parameters. The final score $\mathbf{S} \in \mathbb{R}^{5k}$ is obtained via
\begin{equation}
\small
\mathbf{S} =  \textrm{sparsemax}(\mathbf{S}_{T,I} \circ  \mathbf{S}_{y}).
\label{seq}
\end{equation}
Each score in $\mathbf{S}$ indicates the probability of its corresponding predicate being selected to deduce the head atom $h((T,I),y)$.
Then $\lfloor 5k \times \beta \rfloor$ atoms ranking at the top of $\mathbf{S}$ are selected to complete the clause generation, where $\beta \in (0,1)$ is a hyper-parameter. For instance, if $b_v(v, y)$ and $ b_t(t, y)$ are selected, the clause will become $b_v(v, y) \land b_t(t, y) \Rightarrow h((T,I), y)$. 

\subsection{Clause Evaluation}
\label{Clause-Evaluation}
In Clause Evaluation, we aim to derive the truth value of the head atom for each clause, given body atoms which are instantiated with constants. Specially, given an atom $b_t(t, y)$, its truth value $\mu(b_t(t, y))$ is computed as
\begin{equation}
\small
 \mu(b_t(t, y))=  \textrm{sigmoid}(s[\mathbf{b}_t, \mathbf{p}, \mathbf{b}_t - \mathbf{p}, \mathbf{b}_t \circ \mathbf{p}] \mathbf{W}_{\mu}),
 \label{truthva}
\end{equation}
where $\mathbf{p} \in \mathbb{R}^d$, $\mathbf{p} = \mathbf{o}_t \circ  \mathbf{y}$, and $\mathbf{W}_{\mu} = \mathbf{W}^{4d \times 1}$ is a trainable parameter. Note that $\mathbf{b}_t \in \mathbb{R}^d$, $\mathbf{o}_t \in \mathbb{R}^d$ and $\mathbf{y} \in \mathbb{R}^d$ are representations of $b_t$, $t$, $y$, respectively, $\mathbf{b}_t$ is taken from $\mathbf{B}$, and $s \in \mathbf{S}$  is the importance score corresponding to the atom $b_t(t, y)$.

To obtain the truth value of the head atom, we approximate logic operators $\land$ and $\lor$ using product t-norm, an example of T-Norm (i.e., $T : [0, 1] \times [0, 1] \to [0, 1]$) \cite{tnorm}. Product t-norm defines $T_{\land}(\mu_1, \mu_2) = \mu_1\mu_2$ and  $T_{\lor}(\mu_1, \mu_2) = 1-(1-\mu_1)(1-\mu_2)$, with $\mu_1,\mu_2 \in [0, 1]$ referring to truth values of atoms. With Product t-norm, the truth value of the head atom $\mu(h((T,I), y))$ can be derived as long as the value for each body atom is given. Recall that our GCN model generates representations for each layer $l\in\{0,...,L\}$. Therefore, with logic clauses $b^l_1 \wedge ... \wedge b^l_n \Rightarrow h((T,I),y)$ generated for each layer $l$, we use disjunctive operators to combine clauses across all the layers as 
$(b^0_1 \wedge...) \vee (b^1_1 \wedge...) \vee ... \vee (b^L_1 \wedge ...) \Rightarrow h((T, I), y)$. %And the final truth value of $h((T, I), y)$ can be computed by aggregating certain $\mu^l$ by $\lor$ (e.g., $\mu^0 \lor \ldots \lor \mu^L$).

For the target task of multimodal misinformation detection, given $(T, I)$, we derive truth values $\mu(h((T,I),y))$ for different candidate labels $y$, e.g., $y\in\{NonRumor, Rumor\}$. Then a cross-entropy loss is adopted to train our model in an end-to-end manner which maximizes the truth values for gold labels. During inference, we compare the truth values for both labels and pick the one corresponding to a larger value as the final prediction.

\section{Experiment}
\subsection{Experiment Setup}
We verify the effectiveness of our approach on two public misinformation datasets (\textit{Twitter} and \textit{Weibo}) and further demonstrate its versatility on a sarcasm detection dataset (\textit{Sarcasm}). Three datasets are described as follows:
1) \textit{Twitter} \citep{twitterset} contains 7334 rumors and 5599 non-rumors for training and 564 rumors and 427 non-rumors for testing. 2) \textit{Weibo} \citep{weiboset} includes 3749 rumors and 3783 non-rumors for training and 1000 rumors and 996 non-rumors for testing. 3) \textit{Sarcasm} \citep{sarcasmset} comprises 8642 sarcasm posts and 11174 non-sarcasm posts for training, 959 sarcasm posts and 1451 non-sarcasm posts for validating and 959 sarcasm posts and 1450 non-sarcasm posts for testing. Furthermore, for \textit{Twitter} and \textit{Weibo}, only samples with both text and image are kept, following previous work \citep{twitterset, www22CAFE}. The data pre-processing of \textit{Sarcasm} follows \citet{sarcasmset}.  For all experiments, we set $k=5, g=10$ and $\beta=0.1$. Other details of the implementation and baselines can be found in the appendix. 

\subsection{Overall Performance}
\begin{table*}
\centering
\begin{adjustbox}{max width=0.83\linewidth}
\begin{tabular}{cccc|ccc|ccc}
\hline
\multirow{2}{*}{Dataset} & \multicolumn{2}{c}{\multirow{2}{*}{Method}} & \multirow{2}{*}{Acc} & \multicolumn{3}{c}{Rumor} & \multicolumn{3}{c}{Non Rumor} \\
\cline{5-10}
 & \multicolumn{2}{c}{} &  & Precision  & Recall & F1 Score & Precision  &  Recall & F1 Score\\ 
\cline{1-10}
\multirow{9}{*}{Twitter} & \multirow{2}{*}{Uni-Modal} & Bert \cite{bert} & 0.733 & 0.571 & 0.754 & 0.650 & 0.857 & 0.722 & 0.784 \\ 
&  & ResNet \cite{resnet34} & 0.644 & 0.473 & 0.712 & 0.568 & 0.812 & 0.610 & 0.697 \\ 
\cline{2-10}
& \multirow{8}{*}{Multi-Modal} & Vanilla & 0.784 & 0.669 & 0.683 & 0.676 & 0.843 & 0.834 & 0.838 \\ 

&  & EANN \citep{eann18} & 0.648 & 0.810 & 0.498 & 0.617 & 0.584 & 0.759 & 0.660 \\
&  & MAVE \citep{MVAE19} & 0.745 & 0.745 & 0.719 & 0.758 & 0.689 & 0.777 & 0.730 \\
&  & SAFE \citep{SAFE20} & 0.762 & 0.831 & 0.724 & 0.774 & 0.695 & 0.811 & 0.748 \\
&  & MVNN \citep{MVNN21} & 0.784 & 0.778 & 0.781 & 0.779 & 0.790 & 0.787 & 0.788 \\
&  & CAFE \citep{www22CAFE} & 0.806 & 0.807 & 0.799 & 0.803 & 0.805 & 0.805 & 0.809 \\ 
&  & BMR \citep{BMRAAAI23}& 0.872 & 0.842 & 0.751 & 0.794 & 0.885 & 0.931 & 0.907 \\ 
\cdashline{3-10}
&  & LogicDM~ & \textbf{0.911} & \textbf{0.909} & \textbf{0.816} & \textbf{0.859} & \textbf{0.913} & \textbf{0.958} & \textbf{0.935} \\
\hline
\multirow{9}{*}{Weibo} & \multirow{2}{*}{Uni-Modal} & Bert \cite{bert} & 0.716  & 0.671 & 0.671  & 0.671 & 0.692 & 0.762 &  0.725 \\
 &  & ResNet \cite{resnet34} & 0.678 & 0.701 & 0.638 & 0.668 & 0.658 & 0.720 & 0.688 \\
 \cline{2-10}
 & \multirow{8}{*}{Multi-Modal} & Vanilla & 0.746 & 0.610 & 0.622  & 0.616 & 0.814 & 0.806 & 0.810 \\
 &  & EANN \citep{eann18} & 0.795 & 0.806 & 0.795 & 0.800 & 0.752 & 0.793 & 0.804 \\
 &  & MAVE \citep{MVAE19} & 0.824 & 0.854 & 0.769 & 0.722 & 0.720 & 0.740 & 0.730 \\
 &  & SAFE \citep{SAFE20} & 0.816 & 0.818 & 0.818 & 0.817 & 0.816 & 0.818 & 0.817 \\
 &  & MVNN \citep{MVNN21} & 0.823 & 0.858 & 0.801 & 0.828 & 0.787 & 0.848 & 0.816 \\
 &  & CAFE \citep{www22CAFE} & 0.840 & 0.855 & 0.830 & 0.842 & 0.825 & 0.851 & 0.837 \\ 
&  & BMR \citep{BMRAAAI23} & 0.831 & 0.831 & 0.838 & 0.834 & 0.831 & 0.824 & 0.827 \\ 
\cdashline{3-10}
 &  & LogicDM~ & \textbf{0.852} &\textbf{0.862} & \textbf{0.845} & \textbf{0.853} & \textbf{0.843} & \textbf{0.859} & \textbf{0.851} \\
 \hline
 \end{tabular}
\end{adjustbox}
\caption{Comparison results for multimdoal misinformation detection on Twitter and Weibo datasets.}
\vspace{-15pt}
\label{misinformation}
\end{table*}

\begin{table}
\centering
  \begin{adjustbox}{max width=1.0\columnwidth}
  \begin{tabular}{cc|cccc}
    \hline
    \multicolumn{2}{c}{Model} & Acc & P  & R & F1 \\
    \hline
    \multirow{2}*{Uni-Modal}
    & BERT \cite{bert} & 0.839 & 0.787 & 0.823 & 0.802\\
    & ViT \citep{iclr21vit} & 0.678 & 0.579 & 0.701 & 0.634\\
    \hline
    \multirow{6}*{multimodal}&$\textrm{HFM}$ \citep{sarcasmset} & 0.834 & 0.766 & 0.842 & 0.802\\
    & $\textrm{D\&R Net}$ \citep{DRnet} & 0.840 & 0.780 & 0.834 & 0.806\\
    & $\textrm{Att-BERT}$ \citep{attbert} & 0.861 & 0.809 & 0.851 & 0.829\\
    & $\textrm{InCrossMGs}$ \citep{IncrossMGsacl22}  & 0.861 & 0.814 & 0.844 & 0.828\\
    & $\textrm{HCM}$ \citep{HCMemnlp22}  & 0.874 & 0.818 & \textbf{0.865} & 0.841\\
    \cdashline{2-6}
    & $\textrm{LogicDM}$&\textbf{0.881}&\textbf{0.857}& 0.850 &\textbf{0.853}\\
  \hline
\end{tabular}
\end{adjustbox}
\caption{Comparison results for mutlimodal sarcasm detection on Sarcasm dataset.}
\vspace{-10pt}
\label{sarcasm}
\end{table}
Table \ref{misinformation} and Table \ref{sarcasm} present comparison results for multimodal misinformation detection and sarcasm detection tasks against popular baselines. Despite well-recognized tradeoffs between performance and model interpretability \cite{nsaiijcai20}, both tables indicate our proposed \textbf{LogicDM} consistently surpasses existing state-of-art 
methods in terms of both Accuracy and F1 Score. Especially our model brings 3.9\% and 1.2\% improvements based on accuracy over state-of-art \textbf{BMR} on \textit{Twitter} and \textbf{CAFE} on \textit{Weibo}. Moreover, our model demonstrates superior Precision than other baselines on \textit{Sarcasm}. Such results verify the advantage of the integration of logical reasoning and neural network. We conjecture that logic components may motivate our model to learn useful rules instead of overfitting to noise. In addition, it is also worth mentioning that there is a difference in performance between Rumor and Non Rumor on \textit{Twiiter}, which may be due to unbalanced proportions within the training set. 

Furthermore, it is observed that multi-modality based methods generally outperform uni-modality based methods, suggesting that text and image can provide complementary information to enhance detection performance. In addition, \textbf{CAFE} and \textbf{BMR} can estimate the importance of different modalities to adaptively aggregate unimodal representations by ambiguity measure component and multi-view learning, thus, showing better performance than simple fusion or concatenation. In contrast, our model achieves this goal by softly choosing predicates to induce logic clauses when taking into consideration the logic relationship among these predicates.

\subsection{Interpretation Study}
\begin{figure*}[t]
\centering
\begin{subfigure}{0.235\textwidth}
\centering
\includegraphics[width=\linewidth]{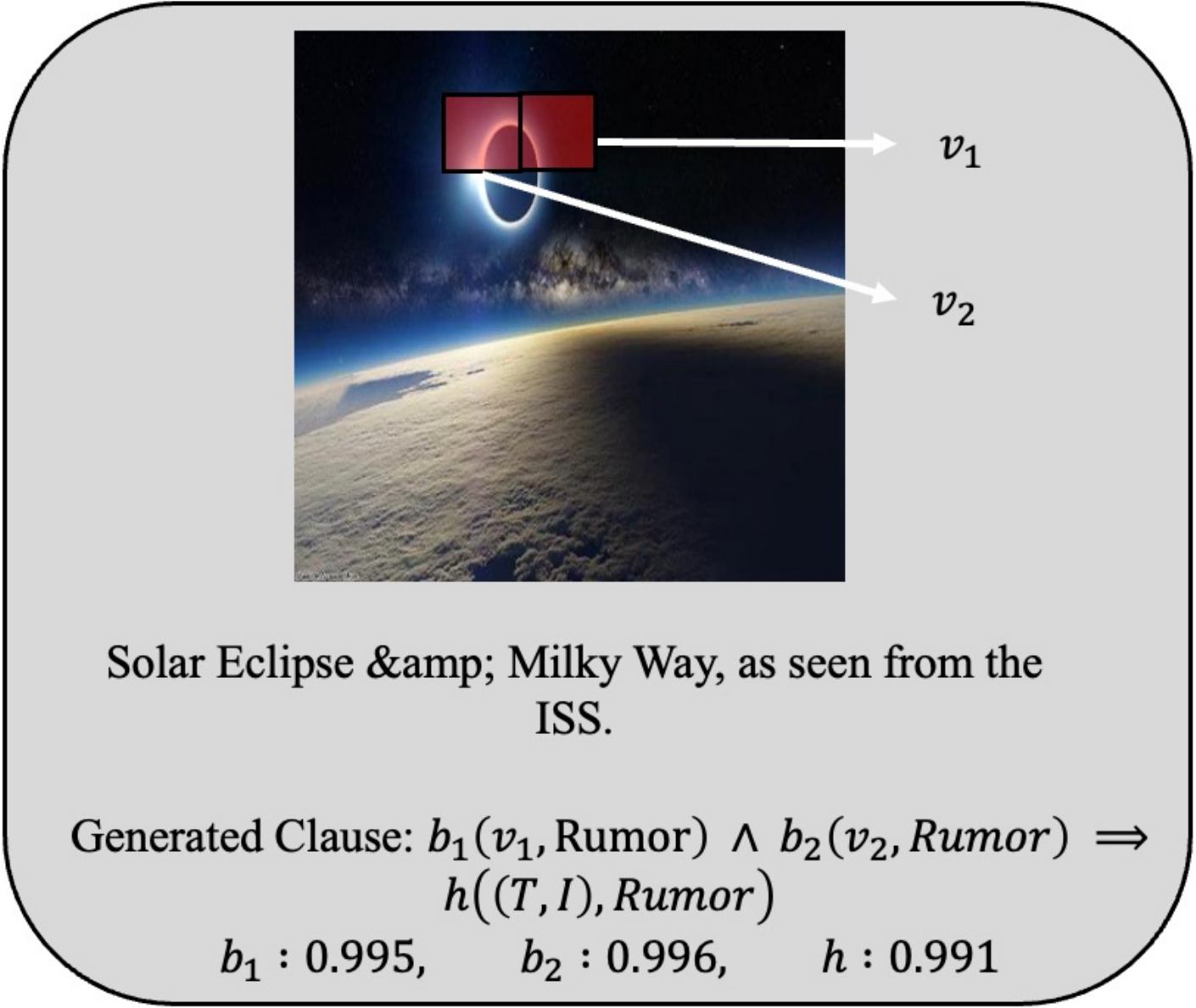}
\caption{Rumor in Twitter}
\label{twitter_smaple_1}
\end{subfigure}
\hfill
\begin{subfigure}{0.235\textwidth}
\includegraphics[width=\linewidth]{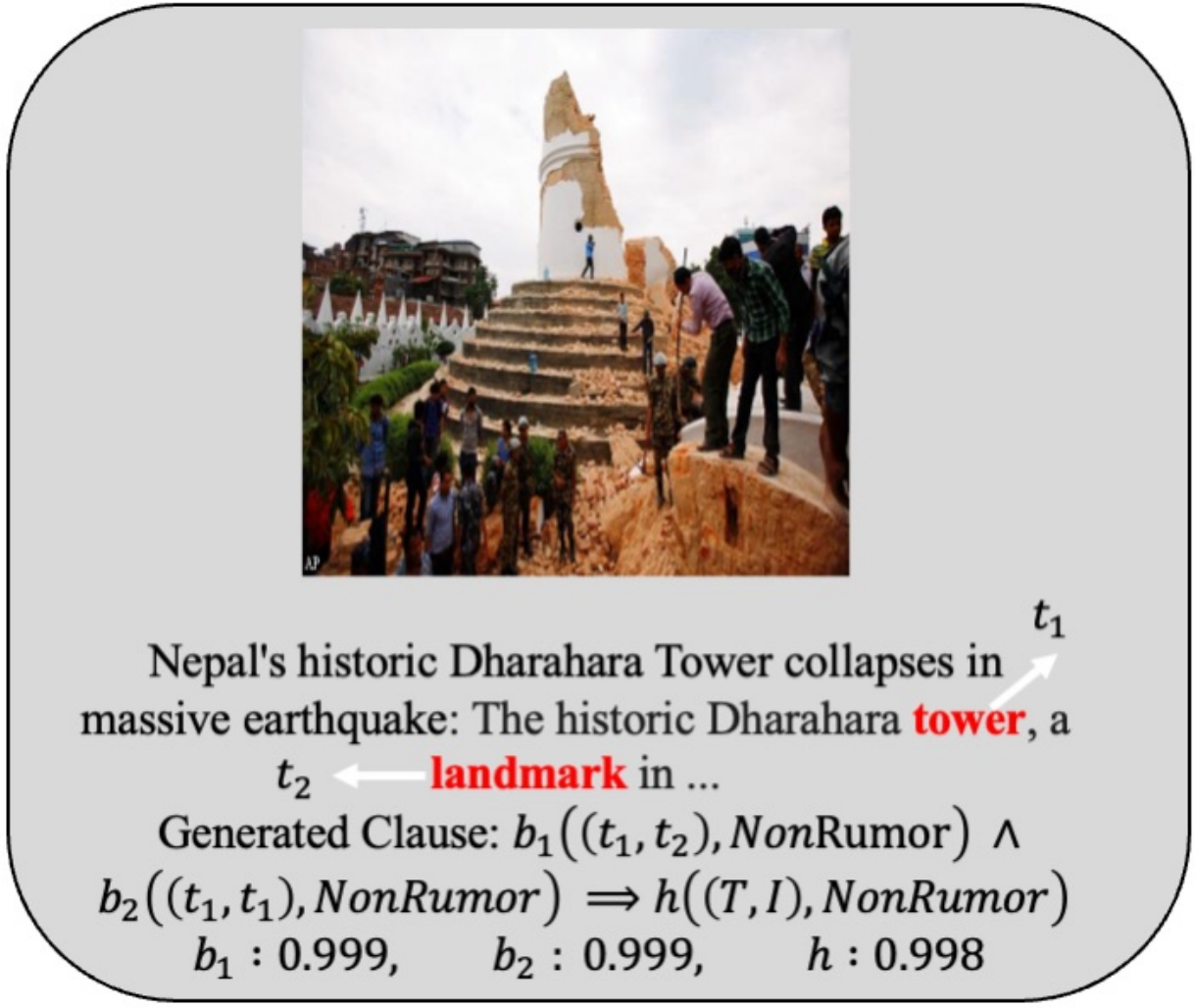} 
\centering
\caption{Non Rumor in Twitter}
\label{twitter_smaple_2}
\end{subfigure}
\hfill
\begin{subfigure}{0.235\textwidth}
\includegraphics[width=\linewidth]{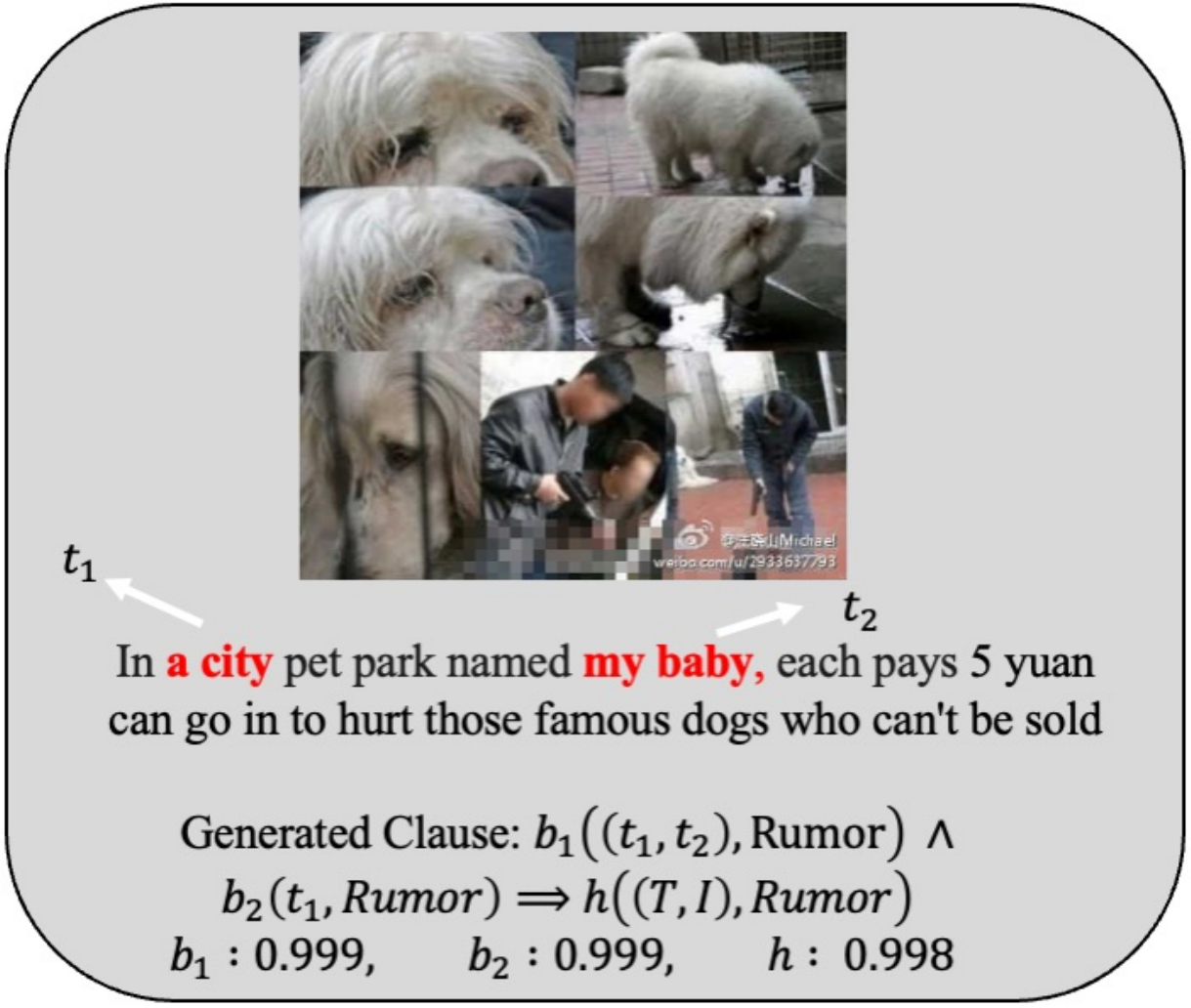} 
\centering
\caption{Rumor in Weibo}
\label{weibo_smaple_1}
\end{subfigure}
\hfill
\begin{subfigure}{0.235\textwidth}
\includegraphics[width=\linewidth]{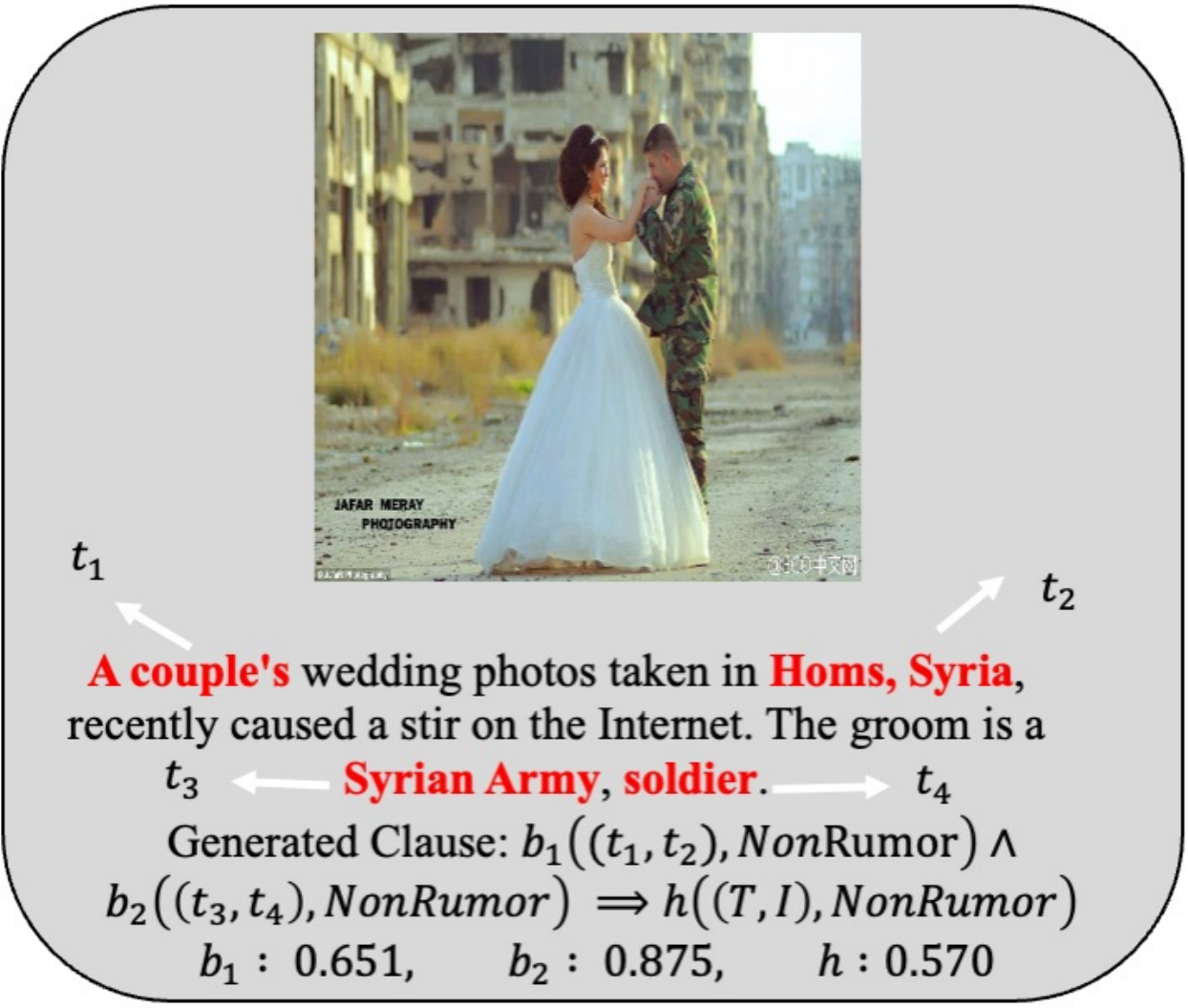} 
\centering
\caption{Non Rumor in Weibo}
\label{weibo_smaple_2}
\end{subfigure}
\caption{Examples of derived clauses and related constants. For (c) and (d), we translate the text from Chinese to English.}
\vspace{-15pt}
\label{interpretationfigure}
\end{figure*}
To illustrate the interpretability of our proposed framework \textbf{LogicDM}, we visualize the learned rules in Fig.~\ref{interpretationfigure}. Despite the complicated text-image input, it is evident that our model can explicitly locate highly correlated content as constants for "where" and softly choose suitable meta-predicates for "how". For example, as shown in Fig.~\ref{weibo_smaple_1}, objects "\textit{a city}" and "\textit{my baby}" are selected to instantiate $b_1$ (i.e., $b_{t,t}$) and $b_2$ (i.e., $b_{t}$) where both predicates implicate that samples with indefinite pronouns are more likely to be rumors. By comparison, samples of proper nouns can usually be detected as non-rumors because of their more realistic description, as seen in Fig.~\ref{weibo_smaple_2}. Moreover, the derived explanation can provide supplementary insights and knowledge previously unknown to practitioners. For example, as seen from Fig.~\ref{twitter_smaple_1}, the logic reasoning based on two visual patches, $b_1$, $b_2$ (i.e., both are $b_{v}$) implies that these areas are hand-crafted\footnote{\url{https://phogotraphy.com/2015/03/20/iss-fake-photo/}} (i.e., produced by Photoshop), which is difficult to be discriminated by human-beings.

Furthermore, our model can mitigate the trust problem of AI systems according to further analyzing derived clauses. For instance, although the nonrumor in Fig.~\ref{twitter_smaple_2} is identified accurately, it may not be sufficiently convincing based on only "\textit{tower}", "\textit{landmark}" and relevant predicates $b_1$, $b_2$ (i.e., both belongs to $b_{t,t}$). In other words, the decision result may not be reliable in this case. The interpretability of the model allows for further understanding of the decision-making process, thus increasing the reliability and trustworthiness of the system.

\subsection{Ablation Study}
\label{Ablation-Study}
In the ablation study, we conduct experiments to analyze the impact of different parameters for performance, including the number of correlations $g$ and rate $\beta$ in Sec.~\ref{Clause-Generation} as well as selected iterations $l$ in  Sec. \ref{Clause-Evaluation}. For illustration, we report the precision, recall, F1 Score of rumor and accuracy on \textit{Twitter} and \textit{Weibo} datasets.
\begin{figure}[t]
\centering
\begin{subfigure}{0.22\textwidth}
\centering
\includegraphics[width=\linewidth]{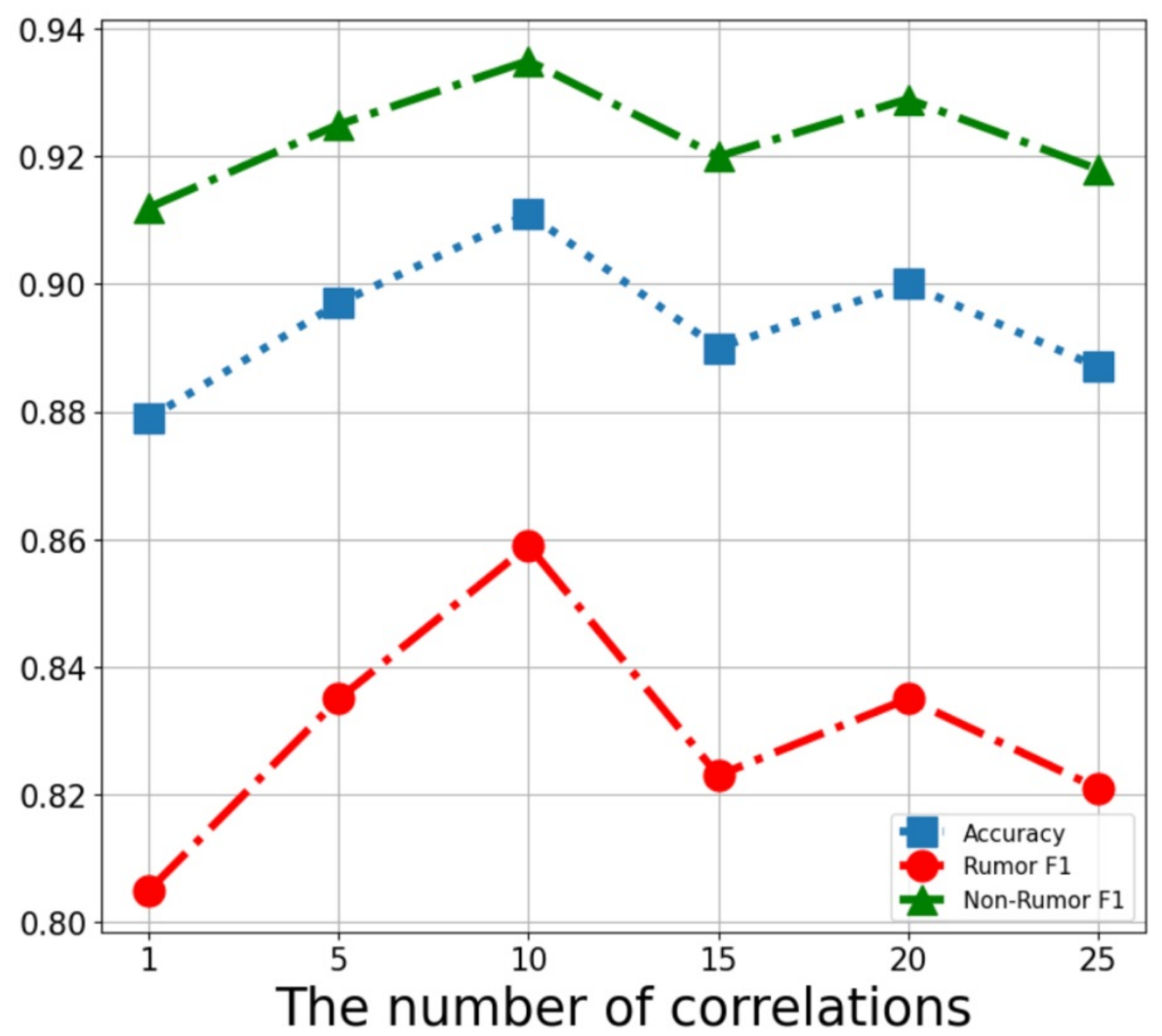}
\caption{Twitter}
\end{subfigure}
\hfill
\begin{subfigure}{0.22\textwidth}
\includegraphics[width=\linewidth]{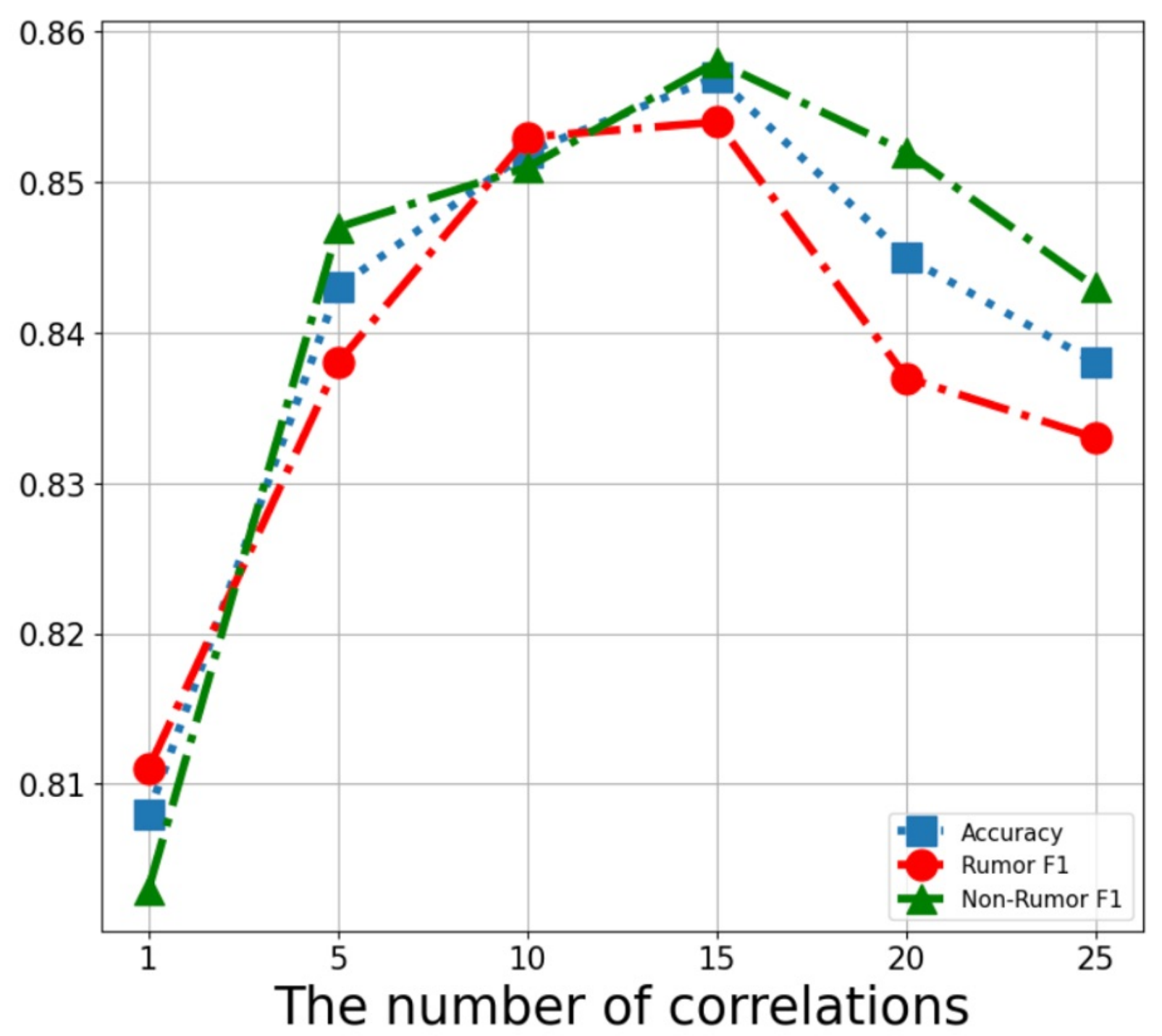} 
\centering
\caption{Weibo}
\end{subfigure}
\caption{The influence of the number of correlations $g$ for dynamic predicate representation.}
\vspace{-10pt}
\label{clues}
\end{figure}

\noindent\textbf{Impact of Number of Correlations.} In order to effectively deal with the diverse online misinformation, we propose to adaptively represent predicates through their corresponding correlation sets in Clause Generation. As seen in Fig.~\ref{clues}, the influence of varying numbers of correlations (i.e., $g$) on performance reveals that the results dramatically increase as $g$ increases and then gradually decrease after reaching a peak (e.g., $10$ for the \textit{Twitter} dataset and $15$ for the \textit{Weibo} dataset). These results validate the effectiveness of dynamic predicate embedding mechanism and suggest that the optimal number of correlations depends on the complexity of specific scenarios. However, it should be noted that our model can be tolerant of an excessive number of correlations without significantly impacting performance.

\begin{figure}[t]
\centering
\begin{subfigure}{0.22\textwidth}
\centering
\includegraphics[width=\linewidth]{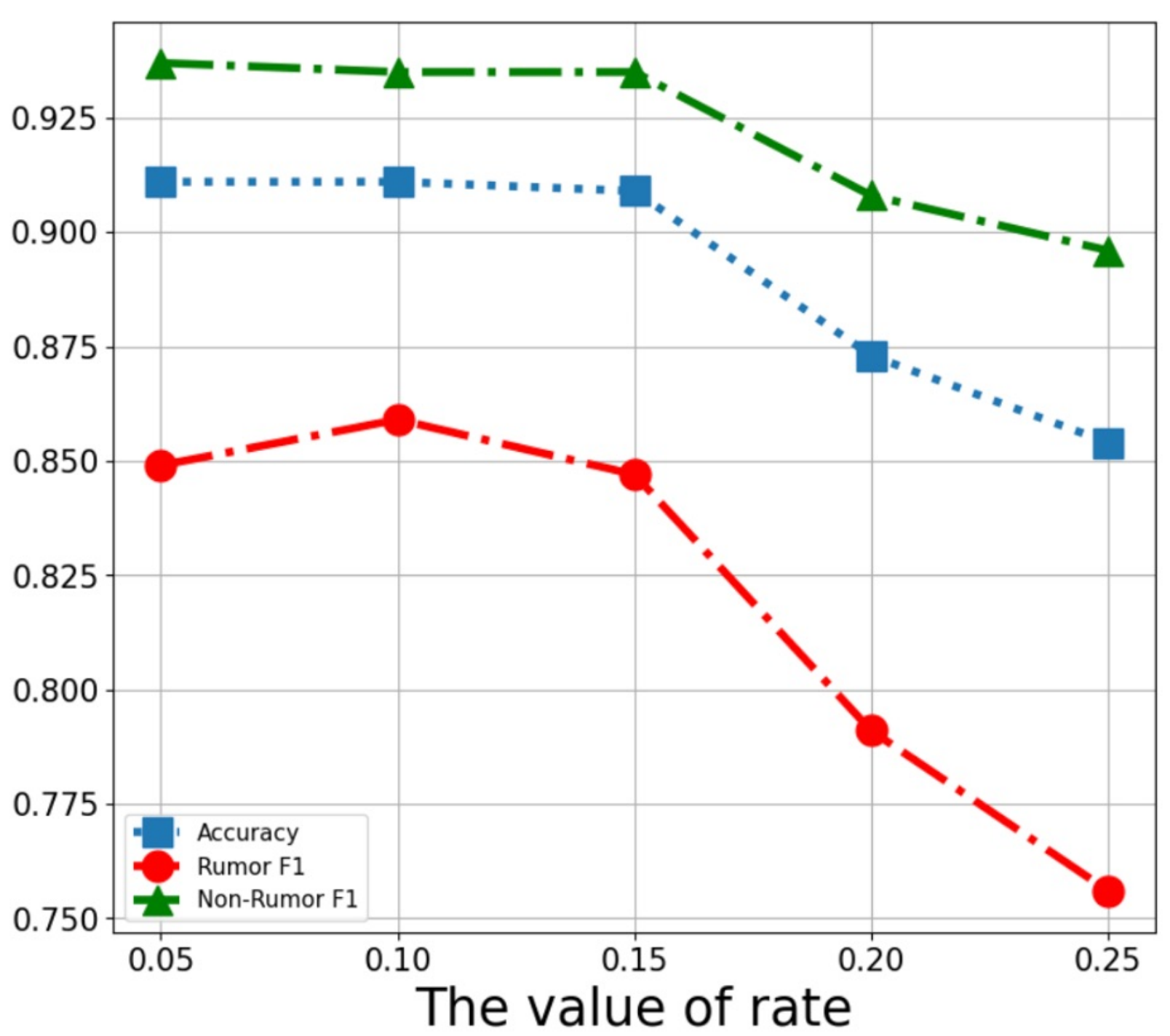}
\caption{Twitter}
\end{subfigure}
\hfill
\begin{subfigure}{0.22\textwidth}
\includegraphics[width=\linewidth]{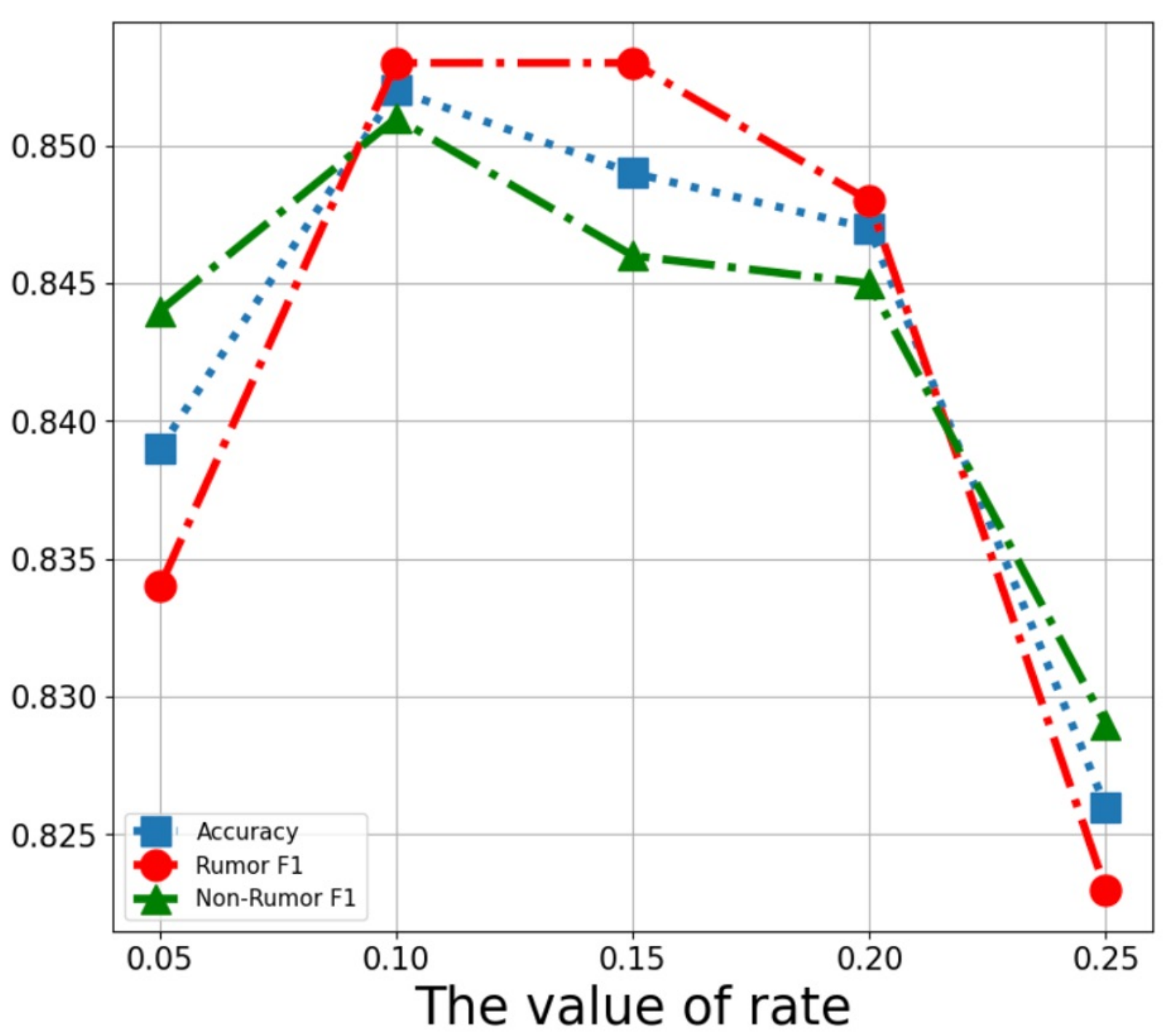} 
\centering
\caption{Weibo}
\end{subfigure}
\caption{The influence of rate $\beta$ for logic clause generation.}
\vspace{-15pt}
\label{rate}
\end{figure}

\noindent\textbf{Impact of Logic Clause Length.} In Clause Generation, we deduce the logic clause of a fixed length by adjusting rate $\beta$. As illustrated in Fig.~\ref{rate}, it is evident that the performance drops significantly as $\beta$ increases from 0.15. This observation can be attributed to two possible reasons: 1) Product t-norm may result in exponential decay when the number of atoms in the clause grows, leading to decreased stability, as previously reported in literature \citep{wenyaacl22multihopreasoning}. 2) Including redundant logic atoms may inevitably introduce noise and negatively impact performance. These findings suggest that a moderate $\beta$ is optimal for clause generation.

\noindent\textbf{Impact of Selected Iterations.} In Clause Evaluation, we obtain the final truth value of head atom $h((T, I), a)$ by selectively aggregating clauses produced at different iterations of GCN based on disjunction operator $\lor$. Table \ref{dis-final-clause} compares various ways for computing $\mu(h((T, I), a))$, revealing that our model achieves the best performance when $l=2$ while yielding the worst performance when $l=0$. Such results highlight the importance of capturing intra-modal and inter-modal interactions of multimodal input through multi-layer GCN for our task. 

Furthermore, it is observed that disjunctive combination  clauses perform more robustly than non-disjunctive combination clauses on \textit{Weibo}, potentially due to the logic-based fusion of information at different iterations. These results provide insights into the importance of incorporating multiple iterations in clauses for better performance in some cases.

\begin{table}
\centering
  \begin{adjustbox}{max width=1.0\columnwidth}
\begin{tabular}{c|ccc|ccc}
\hline
\multirow{2}{*}{selected iteration $l$} & \multicolumn{3}{c}{Twitter} & \multicolumn{3}{c}{Weibo} \\
\cline{2-7}
 & Accuracy &Rumor F1 & Non-Romor F1 & Accuracy & Rumor F1 & Non-Romor F1 \\ 
 \hline
$l \in \{0\}$ & 0.745 & 0.638 & 0.799 & 0.825 & 0.826 & 0.824 \\ 
$l \in \{1\}$ & 0.882 & 0.821 & 0.912 & 0.840 & 0.837 & 0.843 \\
$l \in \{2\}$ & \textbf{0.911} & \textbf{0.859} & \textbf{0.935} & \textbf{0.852} & \textbf{0.853} & \textbf{0.851} \\ 
$l \in \{0, 1\}$ & 0.847 & 0.762 & 0.887 & 0.847 & 0.848 & 0.846 \\ 
$l \in \{1, 2\}$ & 0.902 & 0.842 & 0.928 & 0.841 & 0.832 & 0.849 \\ 
$l \in \{0, 1, 2\}$ & 0.842 & 0.742 & 0.886 & 0.847 & 0.843 & 0.850 \\
\hline
\end{tabular}
\end{adjustbox}
\caption{The influence of selected iterations for clause evaluation. $l\in \{0\}$, $l \in\{1\}$, $l\in\{2\}$ are non-disjunctive combination clauses and the others are disjunctive combination clauses. For example, when $l\in \{0\}$, $h((T,I), a)=(b^0_1 \wedge...)$ and when $l\in \{0, 1\}$, $h((T,I), a)=(b^0_1 \wedge...) \lor (b^1_1 \wedge...)$.}
\label{dis-final-clause}
\vspace{-15pt}
\end{table}

\section{Conclusion}
We propose an interpretable multimodal misinformation detection model \textbf{LogicDM} based on neural-symbolic AI. We predefine five meta-predicates and relevant variables evolved from corresponding misinformation detection perspectives.  And we propose to dynamically represent these predicates by fusion of multiple correlations to cover diversified online information. Moreover, we differentiate reasoning process to smoothly select predicates and cross-modal objects to derive and evaluate explainable logic clauses automatically. Extensive experiments on misinformation detection task demonstrate the effectiveness of our approach and external experiments on sarcasm detection task reveal the versatility.

\section*{Limitations}
Our work has two limitations that may impact the generalization ability of our proposed framework. Firstly, in the Clause Generation section (Sec. \ref{Clause-Generation}), we deduce logic clauses involving a fixed number of atoms, represented by $\lfloor 5k \times \beta \rfloor$, rather than variable length for each iteration of GCN. While this approach has demonstrated superior performance on the multimodal misinformation detection and sarcasm detection tasks, it may harm the generalization of our framework to more complex multimodal misinformation tasks, such as the detection of fake news that involves various modalities, including social networks, text, user responses, images and videos, as discussed in \citep{fakenewssurvey1, fakenewssurveycoling22}. Secondly, in our work, the incorporation of logic into the neural network relies on the use of product t-norm to differentiate logic operators (i.e., $\land$ and $\lor$). However, as shown in the Ablation Study (Sec. \ref{Ablation-Study}), product t-norm may lead to vanishing gradients with the increase of logic atoms during the training stage, which may limit the ability of our proposed framework to handle more sophisticated scenarios. We plan to address these limitations in future research.

\section*{Ethics Statement}
This paper complies with the ACM Code of Ethics and Professional Conduct. Firstly, our adopted datasets do not contain sensitive private information and will not harm society. Secondly, we especially cite relevant papers and sources of pre-trained models and toolkits exploited by this work as detailed as possible. Moreover, our code will be released based on the licenses of any used
artifacts. At last, our proposed multimodal misinformation detection approach will contribute to protecting human beings from the detrimental and unordered online environment with more trustworthy interpretations. 

\section*{ACKNOWLEDGEMENT}
This work was supported in part by CityU Teaching Start-up Grant 6000801, CityU New Research Initiatives/Infrastructure Support from Central (APRC 9610528), the Research Grant Council (RGC) of Hong Kong through Early Career Scheme (ECS) under the Grant 21200522 and Hong Kong Innovation and Technology Commission (InnoHK Project CIMDA).

\newpage
\bibliographystyle{acl_natbib}
\bibliography{ref}

\appendix
\section{Implementation}
 In Feature Extraction, we set $d=200$ and employ pretrained Bert (i.e., bert-base-uncased\footnote{\url{https://huggingface.co/bert-base-uncased}} for \textit{Twitter} and \textit{Sarcasm} and bert-base-chinese\footnote{\url{https://huggingface.co/bert-base-chinese}} for \textit{Weibo}) with one-layer LSTM as textual encoder to extract 200-dimension textual features. For visual modality, we divide the $224 \times 224$ image into $32 \times 32$ patches (i.e., $r=49$, $z=7$). We utilize ResNet34\footnote{\url{https://pytorch.org/vision/main/models/generated/torchvision.models.resnet34}} as  visual backbone for \textit{Twitter} and \textit{Weibo}, following \citep{www22CAFE} and ViT\footnote{\url{https://github.com/lukemelas/PyTorch-Pretrained-ViT}} for \textit{Sarcasm}, following \citep{HCMemnlp22}. The extracted visual features are subsequently mapped to the same dimension as textual features. In Cross-modal Objects Generation, we apply two-layer GCN (i.e., $L=2$) to generate high-level representations of textual tokens and visual patches and then $k=5$ to filter out five candidate objects for each meta-predicate. In Clause Generation, we set the number of correlations $g=10$ and $\beta=0.1$ to derive explainable logic clauses of length $\lfloor 5k\times\beta\rfloor$. At last, we set $h((T, I), a) = b_0^2 \wedge...\wedge b_{\lfloor 5k\times\beta\rfloor-1}^2$ (i.e., $l \in \{2\}$) to obtain the truth value of the target atom in Clause Evaluation. The number of parameters of our model is 4601019 without taking parameters of Bert and the visual backbone neural network (i.e., ResNet and ViT) into account. 
 
During model training, we set the batch size to 32, the epoch number to 20 and exploit Adam optimizer for three sets. Additionally, we adopt an initial learning rate of 0.0001 and a weight decay of 0.0005 for \textit{Twitter} and \textit{Weibo} and 0.00002 and 0.0005 for \textit{Sarcasm}. Moreover, early stopping  strategy is used to avoid overfitting. And we run our experiments on four NVIDIA 3090Ti GPUs.  

For model evaluation, in accordance with prior research \citep{www22CAFE}, we report Accuracy, and Precision, Recall, F1 Score for rumor and non rumor on \textit{Twitter} and \textit{Weibo}, while Accuracy, and Precision, Recall, F1 Score for sarcasm posts on \textit{Sarcasm}. 

\section{Baseline Models}
To comprehensively evaluate our proposed method \textbf{LogicDM}, we divide the baseline models into two categories: Uni-Modal and Multi-Modal methods. For Uni-Modal baselines, we adopt \textbf{Bert} \cite{bert} where the mean embedding of all tokens is utilized for classification and pretrained visual backbone networks where the feature representation after the final pooling layer is used. Specifically, for the visual backbone model, we adopt \textbf{ResNet } \cite{resnet34} for \textit{Twitter} and \textit{Weibo} datasets as suggested by  \citet{www22CAFE}, and adopt \textbf{ViT} \cite{iclr21vit} for sarcasm detection dataset by following \citet{HCMemnlp22}. 

For Multi-Modal baselines, we utilize different approaches for multimodal misinformation detection and sarcasm detection due to the discrepancy between both tasks. Concretely, for \textit{Twiter} and \textit{Weibo}, we adopt \textbf{Vanilla}, \textbf{EANN} \citep{eann18}, \textbf{MAVE} \citep{MVAE19}, \textbf{SAFE} \citep{SAFE20}, \textbf{MVNN} \citep{MVNN21}, \textbf{CAFE} \citep{www22CAFE}, \textbf{BMR} \citep{BMRAAAI23}. Especially, \textbf{Vanilla} fuses the textual and visual features extracted by corresponding encoders of our proposed \textbf{LogicDM} for classification and we re-implement \textbf{BMR} by using the same Feature Extraction component as our method and removing image pattern branch for a fair comparison. For \textit{Sarcasm}, we utilize \textbf{HFM} \citep{sarcasmset}, $\textbf{D}\&\textbf{R Net}$ \citep{DRnet}, \textbf{Att-BERT} \citep{attbert}, \textbf{InCrossMGs} \citep{IncrossMGsacl22} and \textbf{HCM} \citep{HCMemnlp22}.

\section{Running Example}
\begin{figure}[htb]
\centering
\includegraphics[width=\linewidth]{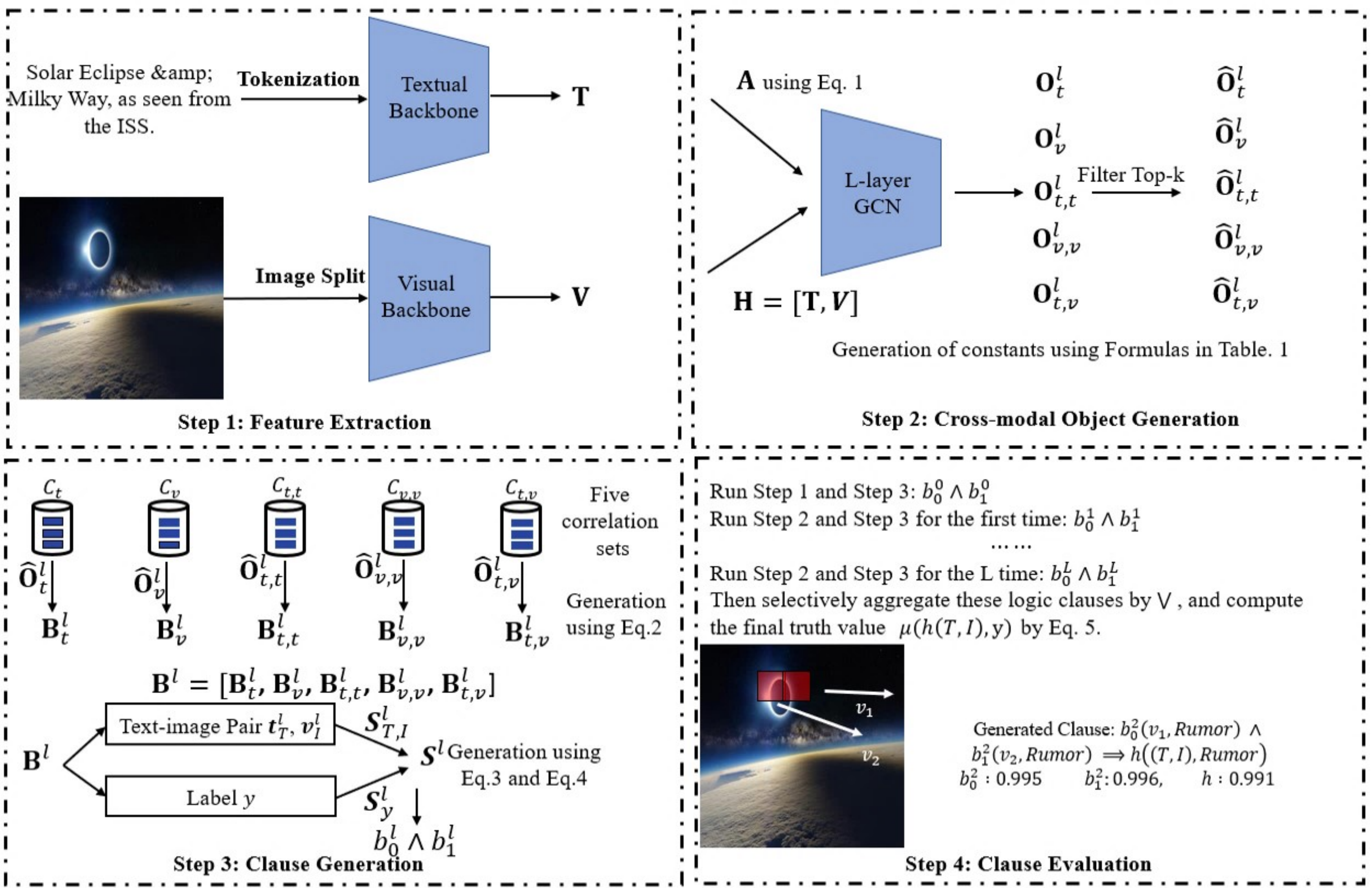}
\caption{The running sample of our proposed \textbf{LogicDM}. In this example, we set $\lfloor 5k \times \beta \rfloor=2$, implying that the derived clauses at each iteration are constituted of two logic atoms and the number of GCN layers is $L=2$.}
\label{runningsample}
\vspace{-10pt}
\end{figure}
To facilitate understanding of the integral reasoning process, we provide an external running example  as depicted in Fig. \ref{runningsample}. The integral reasoning process can be summarized as follows: 1) Given the text-image pair as input, our model first extracts textual features $\mathbf{T}$ and visual features $\mathbf{V}$  using corresponding encoders. 2) These features are exploited to construct a cross-modal graph, denoted by the adjacency matrix $\mathbf{A}$ in Eq. \ref{adja} and node matrix $\mathbf{H} = [\mathbf{T}, \mathbf{V}]$. This graph is fed into an L-layer GCN to conduct cross-modal reasoning. Especially at the iteration \textit{l} of GCN, the output of GCN $\mathbf{H}^l$ is taken to construct cross-modal objects $\mathbf{O}_t^l$, $\mathbf{O}_v^l$, $\mathbf{O}_{t,t}^l$, $\mathbf{O}_{v,v}^l$ and $\mathbf{O}_{t,v}^l$, corresponding to each predicate, using formulas in Table \ref{predicate}. These objects are then refined through a purification process to retain only the most salient ones, denoted as  $\mathbf{\hat{O}}_t^l$, $\mathbf{\hat{O}}_v^l$, $\mathbf{\hat{O}}_{t,t}^l$, $\mathbf{\hat{O}}_{v,v}^l$ and $\mathbf{\hat{O}}^l_{t,v}$, serve as constants to instantiate logic clauses. 3) To derive logic clauses at the iteration \textit{l}, we obtain the predicate representations by weighting the importance of each clue in the corresponding clue set $\mathbf{C}$ for each pair of objects and label y using Eq. \ref{o-predicate}. Then two atoms from $\mathbf{B}^l$ are selected to derive logic clauses $b^l_0 \wedge b^l_1$ based on the importance score $\mathbf{S}^l$ in Eq. \ref{seq}. 4) As each iteration will produce one logic clause, the final logic clause can be deduced by $(b^0_0 \wedge b^0_1) \vee (b^1_0 \wedge b^1_1) \vee ... \vee (b^L_0 \wedge b^L_1) \Rightarrow h((T, I), y)$, of where the truth value can be computed based on Eq. \ref{truthva} and product t-norm. In this example, we only choose $b^2_0(v_1, Rumor) \wedge b^2_1(v_2, Rumor)$ as the final clause. 

\end{document}